\documentclass[%
 aip,
preprint,%
]{revtex4-1}

\usepackage{amsthm,amsmath,amssymb}
\usepackage{graphicx}
\usepackage{dcolumn}
\usepackage{bm}

\usepackage[utf8]{inputenc}
\usepackage[T1]{fontenc}
\usepackage{mathptmx}
\usepackage{mathrsfs} %
\usepackage{float}
\usepackage{subfigure}

\usepackage{textcomp} 
\usepackage{textgreek} 

\begin{document}

\preprint{AIP/123-QED}

\title{Finite volume based film flow and ice accretion models on aircraft wings}

\author{Tong Liu}
  \affiliation{Nuclear Power Institute of China, Chengdu 610213, PR China}
    \affiliation{Department of Fluid Mechanics, Northwestern Polytechnical University, Xi'an, 710072, PR China}
\author{Jinsheng Cai}
  \affiliation{Department of Fluid Mechanics, Northwestern Polytechnical University, Xi'an, 710072, PR China}
\author{Kun Qu}
  \affiliation{Department of Fluid Mechanics, Northwestern Polytechnical University, Xi'an, 710072, PR China}

\author{Shucheng Pan}
  \email{shucheng.pan@nwpu.edu.cn.}
  \affiliation{Department of Fluid Mechanics, Northwestern Polytechnical University, Xi'an, 710072, PR China}

\date{\today}

\begin{abstract}
The thin runback water films driven by the gas flow, the pressure gradient and the gravity on the iced aircraft surface are investigated in this paper.
A three-dimensional film flow model based on Finite Volume Method (FVM) and the lubrication theory is proposed to describe the flow.
The depth-averaged velocity of the film is stored in Cartesian coordinates to avoid the appearance of the metric tensors.
The governing equations are discretized in the first layer structured grid cell which is selected as the grids for film flow.
In order to verify this method, comparisons between numerical results and experimental results of ice shapes on NACA 0012 airfoil and GLC-305 swept wing are presented, both showing a good agreement for rime and glaze ice condition.
Overall, this model shows great potential to model ice accretion reasonably under different icing conditions.
Besides, the present method doesn't require analytic metric terms, and can be easily coupled to existing finite volume solvers for logically Cartesian meshes.
\end{abstract}

\maketitle

\section{Introduction}
\label{sec:introduction}

The ice and water film may occur at the windward side of an aircraft when it flies through clouds in which supercooled droplets are suspended.
With the driven force of the shear stress of the air, pressure gradient and gravity, unfrozen water film may run back downward on the curved aircraft surface, freeze, and form what called is glaze ice.
Glaze ice is a typical glossy ice shape with a single or double horns at the leading edge.
Such severe ice accretion would modify aircraft's geometry, degenerate its aerodynamic characteristics, and pose a serious threat to the safety of the aircraft flight.
Numerical simulation provides a low-cost way to study the ice accretion and runback water film flow on the aircraft surface.

Typically, the film thickness on the iced surface is distinctly smaller than its lateral dimension, and there are many other aspects dealing with the thin water film evolution on curved three-dimensional surface, such as shallow water problem, draining, coating and gravity driven film flow.
Generally, complex mathematical model and a mass of computational time are needed to numerically solve the film flow.
Lubrication (or long-wave, also long-scale) theory\cite{benney_long_1966,oron_long-scale_1997} is usually used to address this problem.
With respect to the full Navier–Stokes equations, lubrication model simplifies the film flow problem reasonably and saves the computational cost by solving only a single degenerate nonlinear equation for the film thickness.
To simulate the evolution of the thickness of a film on a general curved substrate, Roy\cite{roy_lubrication_2002} derived a lubrication model expressed in terms of the film thickness and in a coordinate system fitted to the curved substrate.
The effects of the curvature of the substrate, gravity and inertia are included to accurately describe the film flow.
Based on Roy's work, Roberts\cite{roberts_accurate_2006} derived a more comprehensive model of the dynamics of the film, and the model is expressed in terms of the film thickness and the averaged lateral velocity.
The model resolved wide range of physical interactions between the various physical process of inertia, surface tension, gravity and substrate curvature.
Flow on different substrate shapes, including flat, cylindrical, channel and spherical, were simulated to illustrate its wide application. 
Jean-Luc\cite{jean-luc_transport_2006} studied the steady gravity-driven flow of a thin layer of viscous fluid over a curved substrate with topographical variations. Different from Roy's model\cite{roy_lubrication_2002}, the controlling equation is expressed in nonorthogonal coordinates.
Besides, a correction terms is introduced to the mass-conservation equation and vertical velocity to ensure that the kinematic boundary condition at the free surface is satisfied exactly.
Using lubrication theory, Howell\cite{howell_surface-tension-driven_2003} derived the general leading-order equations governing the flow of a thin liquid film over a moving, curved substrate. Furthermore, the effects of the curvature of the substrate are investigated, and three possible distinguished limits are identified.

Aircraft ice accretion is a much more complex situation involving not only film flow but also phase change, heat transfer and interface propagation.
Based on the Lubrication theory\cite{benney_long_1966} and Stefan problem\cite{james_one_1987}, Myers\cite{myers_slowly_2002} first proposed the mathematical model for ice accretion and water flow on the flat plate.
The glaze icing rate is determined by Stefan condition, and the film flow is driven by air shear, gravity, pressure gradient and surface tension.
For film flow on arbitrary three-dimensional surface, a governing equation is derived based on the curvilinear orthogonal coordinates, in which the first and second fundamental metric tensors are used to carry information about the geometry of the curved surface\cite{myers_flow_2002,Myers_mathematical_2004}.
\citet{cao_extension_2016} extended Myers' model for curvilinear nonorthogonal grids systems.
As a result, besides the first and second fundamental metric tensors, second cross fundamental forms of the surface were included to address the influence of the metrics.
Similarly, to simulate the ice accretion on helicopter rotors, Chen\cite{chen_three-dimensional_2018} developed a three-dimensional ice accretion model on body-fitted nonorthogonal curvilinear coordinates.
In this model, besides the gravity force, the centrifugal caused by the rotation of the coordinate was also accounted as body force.

However, film flow models aforementioned share the same disadvantage that metric tensors should be evaluated to derive the governing equations on curvilinear coordinate systems.
The velocity field is expressed in curvilinear coordinate systems, either.
While in a general finite volume method flow solver, the velocity field is usually expressed in Cartesian coordinate system.
Hence it would take a lot effort to implement the film flow module with airflow solver.
To combat this problem, Calhoun\cite{calhoun_logically_2008,calhoun_finite_2009} proposed a finite volume method for solving parabolic equations on logically Cartesian curved surface, and commented that this method didn't require analytic metric terms, showed second order accuracy and can be easily coupled to existing finite volume solvers for logically Cartesian meshes.
Tukovi${\rm \acute{c}}$\cite{tukovic_moving_2012} and Rauter\cite{rauter_faSavage_2018,rauter_finite_2018} developed the finite area method (FAM), and derived the controlling equation of a depth-integrated shallow flow model for granular materials on three-dimensional mildly curved topographies.
The governing equatuons are expressed in three-dimensional Cartesian coordinates, and share the advantage of easy coupling with three-dimensional ambient flow equations.

In this paper, the governing equations for film flow and ice accretion process on arbitrary curved surfaces are derived based on the finite area method, the Lubrication theory, and the Stefan problem.
The governing equations of film flow are expressed in film thickness and depth-averaged velocity.
The velocity of the film flow is expressed in Cartesian coordinate system, therefore the equations are easy to be coupled with the existing FVM based flow solver Exstream\cite{cai_parallel_2006,xu_numerical_2013,xu_flow_2014}.
Besides, the film flow flow is dominated by shear stress, pressure gradient and gravity.
The velocity profile in the film is approximated as a polynomial function with respect to the film thickness, and the depth-averaged velocity of the film can be expressed as a function of the film thickness and the driving forces acting on the film.

The article is constructed as follows.
Section~\ref{sec:film_flow} derives the film flow governing equations based on the lubrication theory and finite volume method, briefly introduces the ice accretion model, and presents the discretization method.
Section~\ref{sec:results} is devoted to the simulated ice accretion results.
Finally, a conclusion will be given in section~\ref{sec:conclusion}.

\section{Mathematical model}
\label{sec:film_flow}

\begin{figure}[htb]
  \centering
  \includegraphics[width=0.5\textwidth]{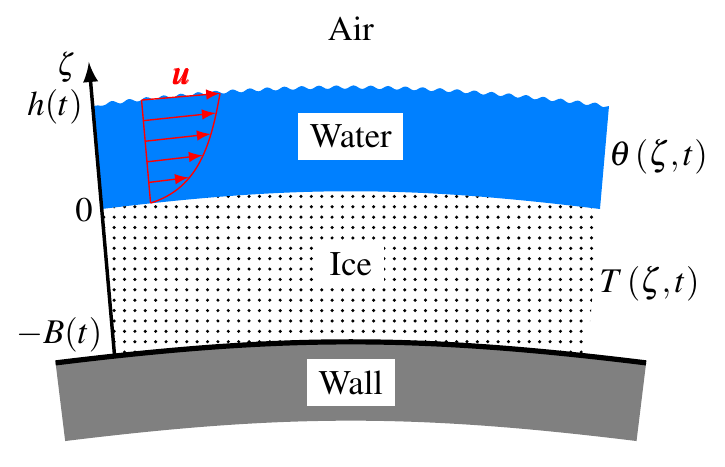}
  \caption{Diagram of the two-dimensional ice and water layers on the substrate.}
  \label{fig:ice_water_layer}
\end{figure}

The problem considered hereby is shown schematically in Fig.~\ref{fig:ice_water_layer}.
Ice accumulates on the curved aircraft surface, and the unfrozen runback water flows over the iced surface driven by the shear stress, pressure gradient and gravity force.
With reference to this figure, $B$ denotes the thickness of the ice in the normal direction to the wall, and $h$ denotes the thickness of the film.
$T$ and $\theta$ are the temperature in the ice and water layers, respectively.
The substrate is denoted as $\mathscr{S}$, and the top surface of the film is denoted as $\mathscr{F}$.
With this article, our aim is to develop a low-dimensional and easy established model for the film flow and icing process on three-dimensional aircraft surface, and hence assumptions are made as follows reasonably:
\begin{itemize}
  \item The film flow is incompressible, namely the density of the film flow is assumed to be constant.
  \item The aspect ratio and the Reynolds number of the thin layer film flow are sufficiently small, which allows the lubrication theory to be used. The the normal velocity component is negligible compared to tangential one, which gives the constraint condition of the film velocity, $\bm{u}\cdot\bm{n}_{wall}=0$ or $\bm{u}=\bm{u}_t$, where $\bm{n}_{wall}$ is the unit normal vector of the wall pointing to the fluid, and $\bm{u}_t$ is the tangential component of the film velocity.
  \item The energy transfer across the film is driven by conduction rather than advection\cite{myers_flow_2002}.
\end{itemize}

\subsection{Film flow model based on the finite volume methods}
\label{sec:film_flow_model_FVM}

\begin{figure}[htb]
  \centering
  \includegraphics[width=0.6\textwidth]{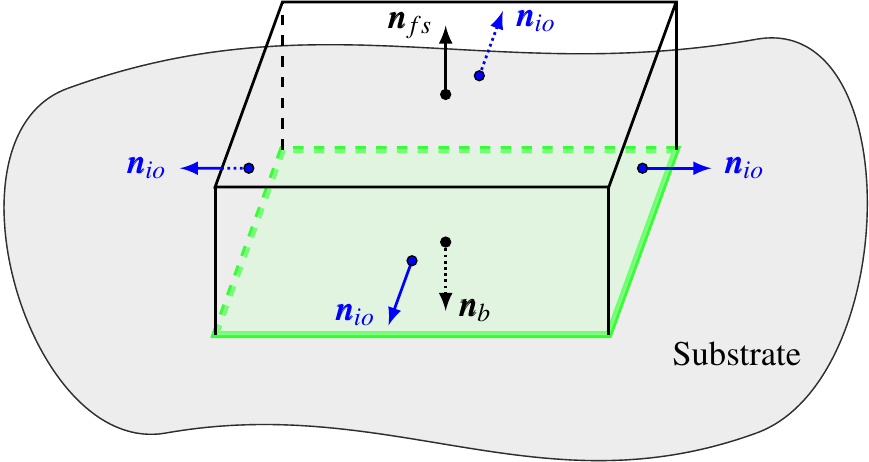}
  \caption{Sketch of the control volume of the film flow.}
  \label{fig:single_control_volume}
\end{figure}

The control volume of the film flow above a patch of the substrate is depicted in Fig.~\ref{fig:single_control_volume}, which is extending across the film layer from $\zeta = 0$ to $\zeta = h$.
Noting that the substrate refers to the ice surface rather than the wall if ice accumulates on the wall.
$\bm{n}_{b}$, $\bm{n}_{fs}$ and $\bm{n}_{io}$ are the outward-pointing unit normal vectors of the faces of the control volume, in which subscript fs, b and io denote the top surface, the bottom surface, and the side surface, respectively.
Because the thickness of the film is very thin, the top surface is assumed to be parallel to the bottom surface, namely $\bm{n}_{wall} = \bm{n}_{fs} = - \bm{n}_{b}$.

\subsubsection{Mass conservation}
\label{sec:mass_conservation}

To model the advection of the film flow on the aircraft surface, we employ integral equations for mass balances written in the conservative forms
\begin{equation}
  \label{eq:mass_conservation_FVM_1}
  \frac{\partial}{\partial t}\int_{V} \rho_w {\rm d}V 
  + \int_{V} \nabla \cdot \left[ \rho_w (\bm{u}-\bm{u}_S) \right] {\rm d}V 
  = \int_{S_{fs}} \dot{m}_{imp} {\rm d}S
  - \int_{S_{b}} \rho_i \frac{\partial B}{\partial t} {\rm d}S
\end{equation}
Where $V$ is the arbitrary control volume for film, and $S$ denotes the surfaces enclosing the control volume $V$, respectively.
$\rho_w$, $\rho_i$, $\bm{u}$, $\bm{u}_S$, $\dot{m}_{imp}$ and $B$ are bulk density of water, bulk density of ice, film velocity, velocity of boundary surface, incoming mass rate due to supercooled droplets impingement and ice thickness respectively.
The second term of right-hand-side of Eq.~\eqref{eq:mass_conservation_FVM_1} denotes mass loss due to ice accretion at the water-ice interface.
Following the assumption that the fluid is incompressible, the density of the film is constant. Applying the Gauss Theorem and extracting the constant density we obtain
\begin{equation}
  \label{eq:mass_conservation_FVM_2}
  \frac{\partial}{\partial t}\int_V {\rm d}V 
  + \oint_S (\bm{u}-\bm{u}_S)\cdot \bm{n} {\rm d}S 
  = \int_{S_{fs}} \frac{\dot{m}_{imp}}{\rho_w} {\rm d}S
  - \int_{S_{b}} \frac{\rho_i}{\rho_w} \frac{\partial B}{\partial t} {\rm d}S
\end{equation}
where $\bm{n}$ is the outward unit normal vector of $S$.
The first term of Eq.~\eqref{eq:mass_conservation_FVM_2} can be transformed into a surface-aligned curvilinear coordinate system:
\begin{equation}
  \label{eq:fist_term_FVM_1}
  \int_V {\rm d}V = \int_{S_b} \int_0^h {\rm det}(\bm{J}){\rm d}\zeta{\rm dS}
\end{equation}
where $\bm{J}$ is the Jacobian matrix induced by coordinate transformation from Cartesian coordinates to curvilinear coordinates.
According to Ref.\cite{thiffeault_transport_2006}, the determinant of matrix $\bm{J}$
\begin{equation}
  \label{eq:determinant}
  {\rm det(\bm{J})}
  = 1 - \kappa\zeta + \mathcal{G}\zeta^2
\end{equation}
where $\kappa$ and $\mathcal{G}$ are the mean curvature and the Gaussian curvature of the surface, respectively.
$\kappa\zeta$ can be estimated as the ratio of flow thickness to curvature radius. The film flow thickness during ice accretion is on the scale of $1.0 \times 10^{-5}{\rm m}$.
Sharp convex and concave are avoided during grid evolution and the maximum grid length along $\xi$ and $\eta$ is usually great than $1.0 \times 10^{-3}{\rm m}$.
Therefore $\kappa\zeta \ll 1$ and hence Eq.~\eqref{eq:fist_term_FVM_1} can be written as
\begin{equation}
  \label{eq:first_term_FVM_2}
  \int_V {\rm d}V
  = \int_{S_b} \int_0^h {\rm d}\zeta {\rm dS}
  = \int_{S_b} h {\rm dS}.
\end{equation}
Similarly, we could obtain
\begin{equation}
  \label{eq:second_term_FVM_1}
  \int_{S_{io}} {\rm d}S = \oint_{L_{io}} \int_0^h {\rm det}(\bm{J}) {\rm d}\zeta {\rm d}L \approx \oint_{L_{io}} \int_0^h {\rm d}\zeta {\rm d}L
\end{equation}
where $L_{io}$ is the side length of the bottom surface.

\begin{figure}[htb]
  \centering
  \includegraphics[width=0.7\textwidth]{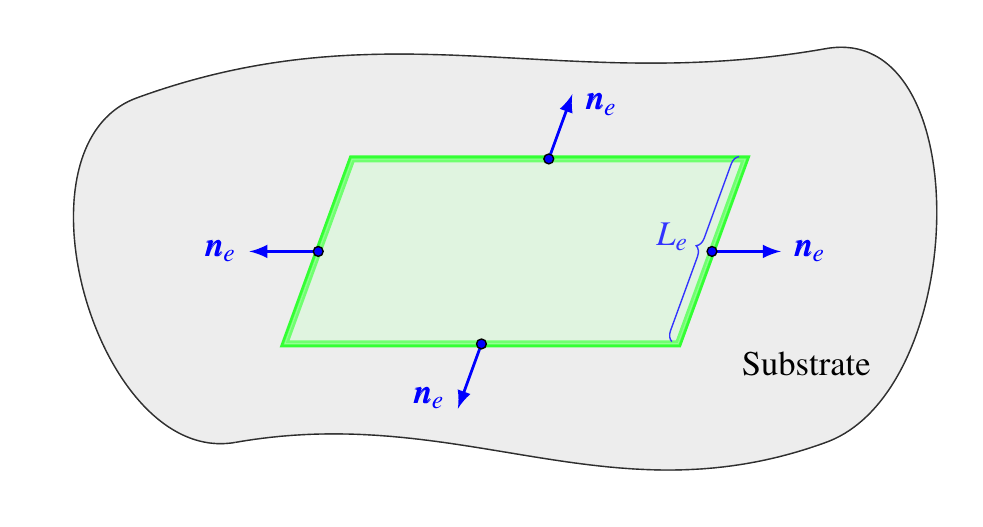}
  \caption{Sketch of the control area of the film flow.}
  \label{fig:diagram_control_area}
\end{figure}

The second term of Eq.~\eqref{eq:mass_conservation_FVM_2} can be split into the integral on each surface:
\begin{equation}
  \label{eq:second_term_1}
  \begin{split}
    &\oint_S (\bm{u}-\bm{u}_S)\cdot \bm{n} {\rm d}S
    = \int_{S_b} (\bm{u}-\bm{u}_b)\cdot \bm{n}_b {\rm d}S \\
    &+ \int_{S_{fs}} (\bm{u}-\bm{u}_{fs})\cdot \bm{n}_{fs} {\rm d}S
    + \int_{S_{io}} (\bm{u}-\bm{u}_{io})\cdot \bm{n}_{io} {\rm d}S
  \end{split}
\end{equation}
The velocity of each surface of the control volume is negligible due to ice grows slowly, namely $\bm{u}_S=\bm{u}_b=\bm{u}_{fs}=\bm{u}_{io}=\bm{0}$.
Besides, the assumptions mentioned before, $\bm{u}\cdot\bm{n}_{wall}=0$ and $\bm{n}_{wall} = \bm{n}_{fs} = - \bm{n}_{b}$, indicate that the first and the second terms of right-hand-side of Eq.~\eqref{eq:second_term_1} are 0.
Combining with Eq.~\eqref{eq:second_term_FVM_1}, the third term of Eq.~\eqref{eq:mass_conservation_FVM_2} cam be written as
\begin{equation}
  \int_{S_{io}} (\bm{u}-\bm{u}_{io})\cdot \bm{n}_{io} {\rm d}S
  = \oint_{L_{io}} \int_0^h \bm{u} \cdot \bm{n}_{io} {\rm d}\zeta {\rm d}L
  = \oint_{L_{io}} \left[ \left( \int_0^h \bm{u} {\rm d}\zeta \right) \cdot \bm{n}_{io} \right] {\rm d}L
\end{equation}
The depth-averaged velocity is defined as
\begin{equation}
  \bar{\bm{u}} = \frac{1}{h} \int_0^h \bm{u} {\rm d}\zeta
\end{equation}
and then Eq.~\eqref{eq:mass_conservation_FVM_2} can be simplified as 
\begin{equation}
\label{eq:second_term_simple}
  \oint_S (\bm{u}-\bm{u}_S)\cdot \bm{n} {\rm d}S
  = \oint_{L_{io}}h\bm{\bar{u}}\cdot\bm{n}_{io}{\rm d}L
\end{equation}

combining with Eq.~\eqref{eq:first_term_FVM_2}, Eq.~\eqref{eq:second_term_simple} and the relation between top surface and bottom surface of the control volume, which reads $S_{fs} = S_{b} {{\rm det} (\bm{J})} \approx S_{b}$, we get the mass conservation equation of the film flow in conservation form 
\begin{equation}
  \label{eq:mass_conservation_FVM_final}
  \frac{\partial}{\partial t} \int_{S_b} h {\rm d}S
  + \oint_{L_{io}}h\bm{\bar{u}}\cdot\bm{n}_{io}{\rm d}L
  = \frac{1}{\rho_w} \int_{S_b} \dot{m}_{imp} {\rm d}S
  - \frac{\rho_i}{\rho_w} \frac{\partial}{\partial t} \int_{S_b} B {\rm d}S
\end{equation}
The control volume $V$ shrinks to the control area as depicted in Fig.~\ref{fig:diagram_control_area}.
$\bm{n}_{e}$ is the outward normal vector of the side of the control area, and is equal to $\bm{n}_{io}$.
The equation above can be rewritten in the differential form for any control area $S_b$\cite{rauter_finite_2018} as
\begin{equation}
  \label{eq:mass_conservation_FVM_final_differential}
  \frac{\partial h}{\partial t}
  + \nabla_s \cdot \left( h \bm{\bar{u}} \right)
  = \frac{\dot{m}_{imp}}{\rho_w} - \frac{\rho_i}{\rho_w} \frac{\partial B}{\partial t}
\end{equation}
where $\nabla_s$ is the Nabla operator on the curves surface.

\subsubsection{Moment conservation}
\label{sec:moment_conservation}

Before deriving the moment conservation equation for the film flow, we introduce the boundary condition related to the film flow first.
The fluid immediately in contact with the bottom substrate $\mathscr{S}$ does not slip along the stationary substrate, namely at the wall the no-slip boundary condition reads $\left. \bm{u} \right|_{\zeta=0} = 0$.
At the film surface $\mathscr{F}$, where $\eta = h$, forces acting on the film surface should be in equilibrium.
The tangential force balance at the air-water interface yields a relation between normal derivative of tangential velocity on the two sides of the water-air interface\cite{tukovic_moving_2012}:
\begin{equation}
  \label{eq:tangential_stress}
  \left. \mu_w \nabla \bm{u} \cdot \bm{n}_{fs} \right|_{\zeta=h} = \left. \mu_a \nabla \bm{u}_{a,t} \cdot \bm{n}_{fs} \right|_{\zeta=h},
\end{equation}
where $\mu_w$ is the dynamic viscosity of water, and $\bm{u}_{a,t}$ is the velocity component of air tangent to the wall.
This equation is equivalent to 
\begin{equation}
  \label{eq:shear_stress_continuty}
  \left. \mu_w \frac{\partial \bm{u}}{\partial n} \right|_{\zeta=h} = \bm{\tau}_a,
\end{equation}
where $\tau_a$ is the air shear stress acting on the film.
The normal force balance at the air-water interface gives
\begin{equation}
  \label{eq:normal_stress}
  \left. p \right|_{\zeta=h} = p_{fs} = p_a + \sigma_{film} \kappa
\end{equation}
where $p_{fs}$ is the pressure of the film at the top surface of the film, $p_a$ is the pressure of the air at the air-water interface, and $\sigma_{film}$ is the surface tension of the film.
Surface tension has little effect on the film velocity, therefore it is negligible during the simulation.

According to the assumptions mentioned before and the lubrication theory, the film flow is treated as an incompressible laminar flow and the convection term can be neglected, therefore the conservation form of the moment equation take the form
\begin{equation}
  \label{eq:moment_conservation_FVM}
  -\int_S p \bm{n} {\rm d}S 
  + \int_V \rho_w \bm{g} {\rm d}V 
  + \int_V \mu_w \nabla^2 \bm{u} {\rm d}V 
  = 0,
\end{equation}
where $p$ is the pressure in the film, and $\bm{g}$ is the gravitational acceleration vector.
Applying the approximation in the Appendix~\ref{app:moment_conservation}, the moment conservation equation can be expressed as integral on the control area
\begin{equation}
  \label{eq:moment_conservation_FAM}
  \begin{split}
    &- \oint_{L_{io}} \bm{n}_{io} h \bar{p} {\rm d}L
    - \int_{S_b} (p_b-p_a) \bm{n}_{wall}  {\rm d}S \\ 
    &+ \int_{S_b} h \rho_w \bm{g} {\rm d}S 
    + \int_{S_{b}}  \bm{\tau}_a {\rm d}S - \int_{S_{b}} \bm{\tau}_b {\rm d}S 
    = 0,
  \end{split}
\end{equation}
Rewritting this equation in differential form, we obtain
\begin{equation}
  \label{eq:differential_moment_equation}
  - \nabla_s \left( h \bar{p} \right) - (p_b-p_a) \bm{n}_{wall}
  + h \rho_w \bm{g}
  + (\bm{\tau}_a - \bm{\tau}_b)
  = 0
\end{equation}
Applying surface normal and surface tangential projection to Eq.~\eqref{eq:differential_moment_equation} yields
\begin{equation}
  p_b = p_a + h \rho_w g_n
\end{equation}
and
\begin{equation}
  \label{eq:normal_moment_equation}
  - \nabla_s \left( h \bar{p} \right)
  + h \rho_w \bm{g}_s
  + (\bm{\tau}_a - \bm{\tau}_b)
  = 0,
\end{equation}
where $\bm{g}_n = \bm{n}_{wall} \cdot \left( \bm{n}_{wall} \cdot \bm{g} \right)$ and $\bm{g}_s = \left( \bm{I} - \bm{n}_{wall} \bigotimes \bm{n}_{wall} \right) \cdot \bm{g}$.
$\bar{p}$ is the depth-averaged pressure of the film, which reads
\begin{equation}
  \bar{p}=\dfrac{p_a+p_b}{2} = p_a + \frac{1}{2} h \rho_w g_n.
\end{equation}

The velocity profile in the water film usually follows a linear\cite{bourgault_development_2000} or polynomial function\cite{myers_extension_2001,myers_flow_2002,wang_thin_2009,cao_extension_2016,chen_three-dimensional_2018} along the film thickness $\zeta$, and the latter leads to
\begin{equation}
  \label{eq:velocity_profile_unknown}
  \bm{u}(\zeta) = \bm{a}\zeta^2 + \bm{b}\zeta + \bm{c},
\end{equation}
where $\bm{a}$, $\bm{b}$ and $\bm{c}$ are the coefficient vectors.
Applying the no-slip boundary condition at the water-substrate interface and the shear stress boundary condition~\eqref{eq:shear_stress_continuty} and integrating the velocity along the film thickness finally yields
\begin{equation}
  \label{eq:averaged_velocity_form_1}
  \bar{\bm{u}}
  = \frac{\left( \bm{\tau}_a + 2 \bm{\tau}_b \right)h}{6 \mu_w}
\end{equation}
The vectors $\bm{\tau}_a$ and $\bm{\tau}_b$ are parallel to the substrate, which makes the film velocity satisfy the constraint condition $\bm{u} \cdot \bm{n}_b = 0$ automatically.
Moreover, combining Eq.~\eqref{eq:normal_moment_equation} and Eq.~\eqref{eq:averaged_velocity_form_1} gives another form of the depth-averaged velocity:
\begin{equation}
  \label{eq:averaged_velocity_form_2}
  \bar{\bm{u}}
  = \frac{h}{2\mu_w} \bm{\tau}_a
  + \frac{h}{3\mu_w} \left[ - \nabla_s \left( h \bar{p} \right) + h \rho_w \bm{g}_s \right]
\end{equation}
It indicates that air shear play a key role in the driving forces of the film flow.
Furthermore, the film flux in Eq.~\eqref{eq:mass_conservation_FVM_final_differential} can be expressed as
\begin{equation}
  \label{eq:film_flux}
  \bm{F}
  = h\bar{\bm{u}}
  = \frac{h^2}{2\mu_w} \bm{\tau}_a
  + \frac{h^2}{3\mu_w} \left[ - \nabla_s \left( h \bar{p} \right) + h \rho_w \bm{g}_s \right]
\end{equation}

\subsubsection{Energy conservation}
\label{sec:energy_conservation}
The temperature at the ice-water interface is freezing temperature of water $T_f$, and the film is very thin. Therefore the depth-averaged temperature of the film is close to $T_f$, hence the assumption is made that there is no heat exchange between adjacent control areas along the wall surface.
Applying the lubrication theory and the approximation aforementioned in a similar way, the energy conservation equation for the film flow on the arbitrary curved surface can be reduced to\cite{myers_slowly_2002,myers_flow_2002}
\begin{equation}
  \label{eq:energy_conservation_FAM}
    \int_V \nabla^2 \theta {\rm d}V = 0.
\end{equation}
The equation above indicates that the temperature gradient in the film normal to the wall is constant, i.e., the temperature profile in the film in linear.

\subsection{Rime and glaze ice accretion model}
\label{sec:rime_and_glaze_ice_accretion_model}
Under rime ice condition, all impinging droplets freeze and there is no film flow on the surface, hence all terms on the left-hand side of Eq.~\eqref{eq:mass_conservation_FVM_final} are identically zero.
The icing rate is simply proportional to the droplet mass rate impinging on the control area, and the control equation can be written in differential form as
\begin{equation}
  \label{eq:rime_ice_rate}
  \frac{\partial B}{\partial t} = \frac{\dot{m}_{imp}}{\rho_i}.
\end{equation}

Under glaze ice condition instead, by applying the energy balance on the ice-water interface\cite{myers_slowly_2002,myers_flow_2002}, the icing rate in differential form can be derived as
\begin{equation}
  \label{eq:glaze_ice_rate}
  \rho_i L_f \frac{\partial B}{\partial t} 
  = k_i \left. \frac{\partial T}{\partial n} \right|_{\zeta=0}
  - k_w \left. \frac{\partial \theta}{\partial n} \right|_{\zeta=0},
\end{equation}
where $L_f$, $k_i$ and $k_w$ are the latent heat of ice accretion, the thermal conductivities of ice and water, respectively.
Eq.~\eqref{eq:glaze_ice_rate} implies that the energy released during the solidification of impinging water is conducted away through the ice and water layers, which is also known as the Stefan condition or phase change condition\cite{james_one_1987}.
The temperature distribution in the ice layer is assumed to be linear\cite{myers_flow_2002,myers_slowly_2002}. Under this assumption, the energy balance~\eqref{eq:glaze_ice_rate} can be expressed in its final form
\begin{equation}
  \label{eq:glaze_ice_rate_detail}
  \rho_i L_f \frac{\partial B}{\partial t} 
  = k_i \frac{T_f-T_s}{B}
  - k_w \frac{ Q_{glaze}^{+} - q_{glaze}^{-} \left(T_f-T_a\right) }{ k_w + hq_{glaze}^{-} },
\end{equation}
where $Q_{glaze}^{+}$ and $q_{glaze}^{-}$ are the heat energy gained and the rate of the heat energy lost during ice accretion, respectively. Details of these two terms are referred to \citet{liu_three-dimensional_2019}.

\subsection{Discretization and solution methods}
\label{sec:discretization}

The ice accretion on the aircraft surface is governed by Eq.~\eqref{eq:mass_conservation_FVM_final_differential}, Eq.~\eqref{eq:normal_moment_equation}, Eq.~\eqref{eq:rime_ice_rate} and Eq.~\eqref{eq:glaze_ice_rate_detail}.
The ice thickness $B$, the film thickness $h$ and the depth-averaged velocity of the film $\bar{\bm{u}}$ are unknown, and now we seek the discretization and solution methods of the governing equations to solve this problem.
Under rime ice condition (dry accretion), the ice thickness increment is solved according to the discretized form of Eq.~\eqref{eq:rime_ice_rate}
\begin{equation}
  \label{eq:rime_ice_rate_discretized}
  \Delta B^n = \frac{\dot{m}_{imp} \Delta t}{\rho_i},
\end{equation}
where $\Delta t$ is the time step employed in the calculation.
Under glaze ice condition (wet accretion), the discretized form of Eq.~\eqref{eq:glaze_ice_rate_detail} gives
\begin{equation}
  \label{eq:glaze_ice_rate_detail_discretized}
  \Delta B^n
  = \frac{\Delta t}{\rho_i L_F}
  \left[
    k_i\frac{T_f-T_s}{B^n}
    -k_w\frac{Q_{glaze}^{+}-q_{glaze}^{-}\left(T_f-T_a\right)}{k_{w}+h^n q_{glaze}^{-}}
  \right].
\end{equation}

The pressure gradient of the film should be solved first to evaluate the depth-averaged velocity and the flux of the film flow explicitly, and applying the Gauss Theorem yields
\begin{equation}
  \label{eq:pressure_gradient_1}
  \nabla_s \left( h \bar{p} \right)
  = \frac{\sum_e \bm{n}_e h_e^n \bar{p}_e L_e}{S_b}
\end{equation}
where $L_e$ is the edge length and $\bm{n}_e$ is the unit normal vectors on lateral surface.
A high-quality body-fitted structured grids with orthogonality of grids ensured near the wall are generated hence $\bm{n}_e$ is equal to $\bm{n}_{io}$ approximatively.
The mass conservation equation~\eqref{eq:mass_conservation_FVM_final} can be discretized as follows to obtain the film thickness increment
\begin{equation}
  \label{eq:film_thickness_discretized}
  \Delta S \frac{\Delta h^n}{\Delta t}
  + \sum_{e} \bm{F}^{n+1} \cdot \bm{n}_{e} L_e
  = \Delta S
  \left(
    \frac{\dot{m}_{imp}}{\rho_w}
    - \frac{\rho_i}{\rho_w} \frac{\Delta B^n}{\Delta t}
  \right)
\end{equation}
The discretization of the convective term is performed with the first-order upwind scheme to smooth the oscillations of the film thickness and the Line Successive Over Relaxation (LSOR) is used to solve the equations on the iced surface.

A solution strategy presented by \citet{myers_flow_2002} is adopted to solve the ice accretion problem.
The type of the ice accretion at each control area is assumed to be wet accretion therefore the ice and film thickness are determined by Eq.~\eqref{eq:glaze_ice_rate_detail_discretized} and Eq.~\eqref{eq:film_thickness_discretized}, which requires that $B^n \neq 0$.
Hence a precursor ice thickness $B_p = 1.0\times10^{-12}\mathrm{m}$ is specified initially to avoid the problem mentioned above.
Besides, a precursor film $h_p$ is introduced to avoid the difficulty associated with the advancing contact line.
In this paper, the aircraft surface is set to be covered with a thin film whose thickness is $1.0\times10^{-12}\mathrm{m}$ initially.
The wet accretion assumption holds in the control area if the new film thickness $h^{n+1}$ is greater than the precursor film thickness $h_p$.
While the wet assumption is invalid if the new film thickness is smaller than the precursor film thickness.
In this case, the new ice thickness increment is calculated by Eq.~\eqref{eq:rime_ice_rate_discretized} and the film thickness is set to $h_p$.

\subsection{Numerical methods}
\label{sec:ice_accretion_model}
\begin{figure}[hb]
  \label{fig:ice_accretion_flowchart}
  \includegraphics[width=0.6\textwidth]{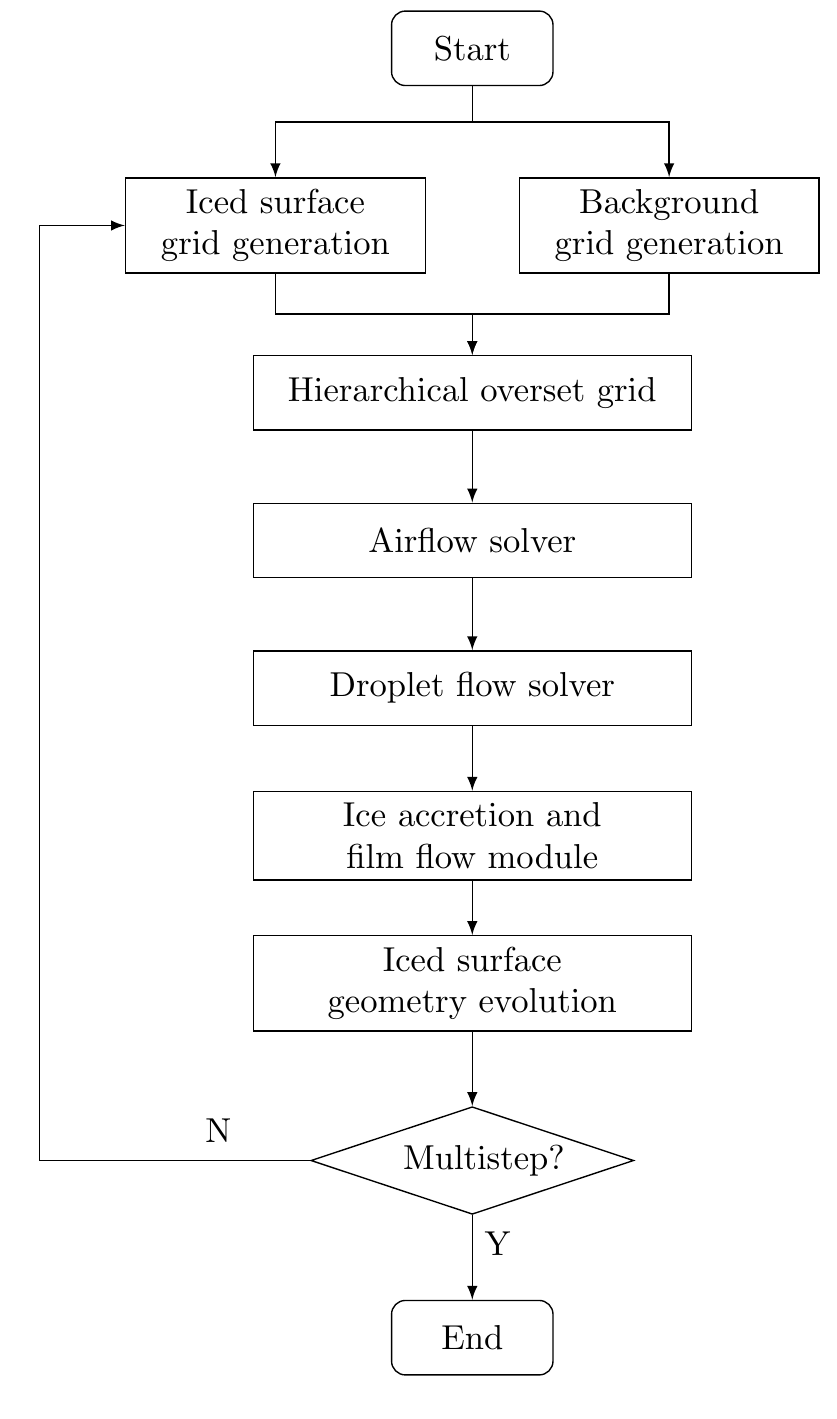}
  \caption{Flowchart of ice accretion.}
\end{figure}

As shown in Fig.~\ref{fig:ice_accretion_flowchart}, Current ice accretion code is composed of several modules, i.e., a flow solver, a droplets flow solver, an ice accretion module with thermodynamic and film flow model, and a mesh module.
The air flow solver called Exstream\cite{xu_numerical_2013,xu_flow_2014,cai_parallel_2006} solves the Reynolds-averaged Navier–Stokes (RANS) equations to evaluate the driving force acting on the film.
The influence of the surface roughness on the convective heat transfer characteristics is taken into account by extending the turbulence model\cite{knopp_new_2009}. 
The droplet flow solver based on the Eulerian method is utilized to obtain the water collection efficiencies on the wall.
The mesh module adopts hierarchical overset grid strategy\cite{cai_parallel_2006} and parabolic grid generation approach to generate high-quality curvilinear body-fitted structured grids on the clean and complex iced geometry.
A quasi-steady multi-step algorithm\cite{verdin_multistep_2009} is employed to simulate the unsteady icing process.
During the unsteady icing process in each single step, the geometry of the iced airfoil changes slowly, implying that the unsteady effects of the air flow can be neglected.
Thus, the parameters of the air flow and droplet flow are kept as constant during the simulation of ice accretion and film flow.
For further information on these modules mentioned above, the reader is referred to \citet{liu_three-dimensional_2019}.

\section{Numerical results and Analysis}
\label{sec:results}

The verification of the film flow and ice accretion models is verified on both two and three dimensional cases where numerical results obtained by current method are compared with available experimental and numerical results.
In addition, the film thickness distribution under different icing conditions are studied.

\subsection{Film flow on NACA 0012 airfoil}
A simulation performed on NACA0012 airfoil presented by \citet{lavoie_modeling_2017} is carried out to verify current film flow model.
The test parameters are the same with that of Run 308 in Table~\ref{tb:naca0012}, except that the angle of attack is 0$^{\circ}$, the temperature is 288.15$\mathrm{K}$ and the exposure time is 4$\mathrm{s}$.

\begin{figure}[hb]
  \label{fig:naca0012_film_thickness_vel}
  \includegraphics[width=1.0\textwidth]{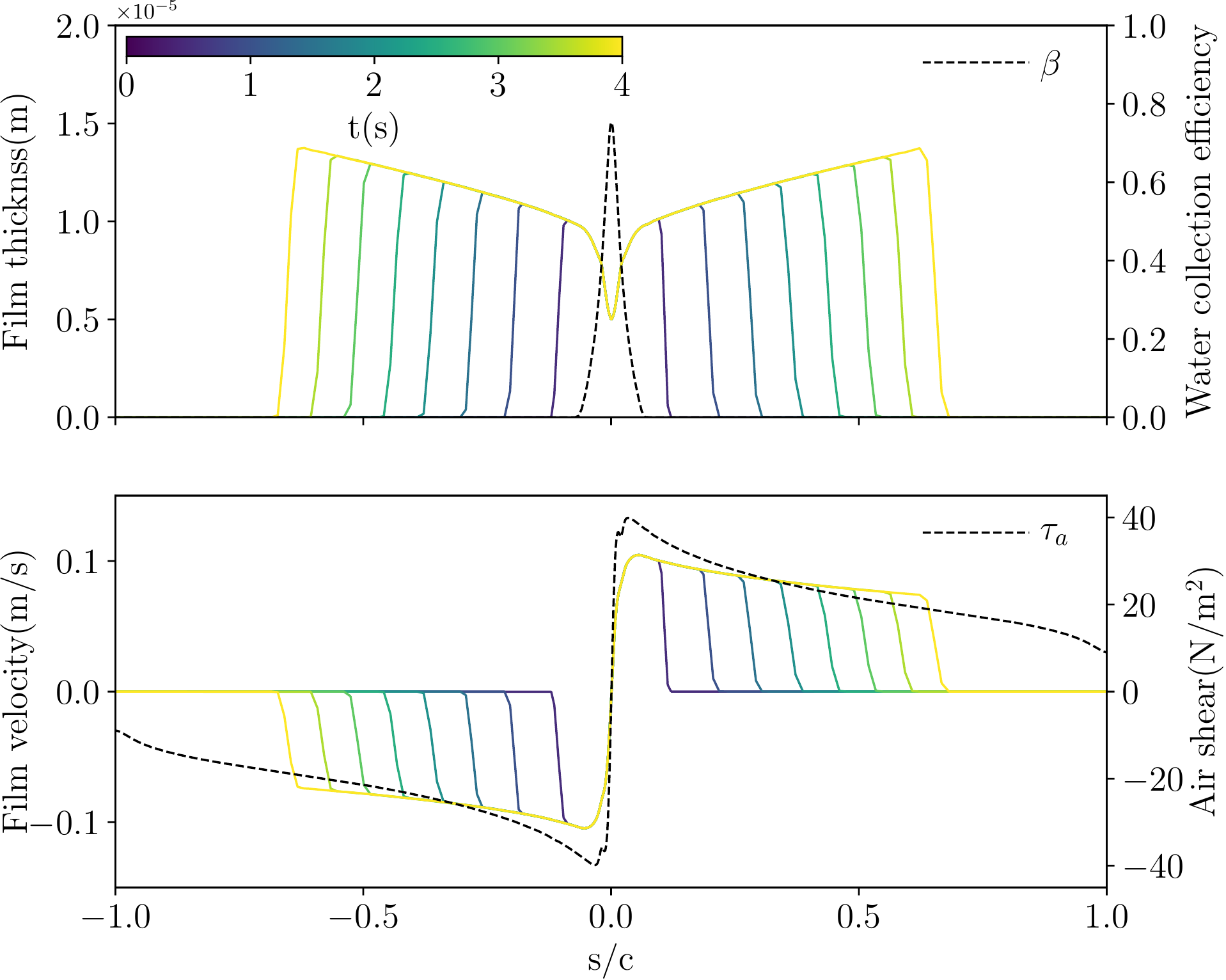}
  \caption{The film flow thickness, the film flow depth-averaged velocity, the water collection efficiency and the air shear stress acting on the film against the non-dimensional curvilinear distances computed clockwise from the trailing edge. The stagnation point locates at $\mathrm{s/c}=0$.}
\end{figure}

Results of the numerical solution are presented in Fig.~\ref{fig:naca0012_film_thickness_vel} for the film thickness evolution, the film depth-averaged velocity evolution, the water collection efficiency and the air shear stress.
The colored solid lines, which are the time series of streamwise profiles plotted every 0.5 $\mathrm{s}$, present the distribution of film thickness and the film velocity, while the black dashed lines present the distribution of the water collection efficiency and the air shear stress acting on the film.
Velocity and air shear stress in the counter-clockwise direction along the airfoil surface is taken as negative values.
Note that the two subfigures share the same X axis.
As we can see, the impinging limits of the droplets locates around the stagnation point and its range is 6.6\% of the surface arc length.
Since the surrounding temperature is warm enough, impinging droplets do not freeze at all and flow all the way from the impinging area to the unimpinging area under the influence of the driving force.
The upper subfigure in Fig.~\ref{fig:naca0012_film_thickness_vel} shows that the film cover about 70\% of the airfoil surface at $\rm{t=4s}$.
The lower subfigure shows that a sharp increase of the film velocity is observed near the stagnation point, which indicates that water around the stagnation point is flow downstream rapidly.
Hence the film thickness around the stagnation point is smaller than that at downstream distinctly even though the water collection efficiency reaches its maximum value at the stagnation point.
Eq.~\eqref{eq:averaged_velocity_form_2} states that the air shear dominates the depth-averaged velocity of the film, thus a decrease of the air shear stress leads to the decrease of the film velocity.
As a result, the film thickness at the advancing front increases over time gradually.

\subsection{Ice accretion on NACA 0012 airfoil}
\label{sec:naca0012}

Wright presented a series of icing results for different icing conditions on different airfoils\cite{wright_validation_1999}.
Experimental results of NASA Lewis Icing Research Tunnel (IRT) and numerical results of LEWICE 2.0 are provided for each case.
In this paper, 9 cases on NACA 0012 airfoil, including rime and glaze ice accretion, are selected to validate current model thoroughly.
The corresponding parameters for these cases are presented in Table~\ref{tb:naca0012}.
The chord length of the airfoil is 0.5334${\rm m}$, and the angle of attack used in these cases is 3.5$^{\circ}$.

\begin{table*}[hb]
  \label{tb:naca0012}
  \caption{Ice conditions for NACA0012 airfoil.}
  \begin{ruledtabular}
    \begin{tabular}{cccccc}
      Case & $U_{\infty}$, $\rm m/s$ & $T_{\infty}$, $\rm K$ & LWC, $\rm g/m^3$ & MVD, $\rm \mu m$ & Time, $\rm min$ \\ \hline
      Run 308 & 102.8 & 262.04 & 1.0  & 20 & 3.85 \\
      Run 316 & 102.8 & 262.04 & 0.55 & 20 & 3.22 \\
      Run 401 & 102.8 & 265.37 & 0.55 & 20 & 7 \\
      Run 403 & 102.8 & 262.04 & 0.55 & 20 & 7 \\
      Run 405 & 102.8 & 250.37 & 0.55 & 20 & 7 \\
      Run 409 & 67.1  & 265.07 & 1.3  & 30 & 6 \\
      Run 421 & 67.1  & 268.40 & 1.0  & 20 & 6 \\
      Run 422 & 67.1  & 266.74 & 1.0  & 20 & 6 \\
      Run 423 & 67.1  & 265.07 & 1.0  & 20 & 6 \\
    \end{tabular}
  \end{ruledtabular}
\end{table*}

\begin{figure*}[hb]
  \label{fig:naca0012_ice_shapes}
  \includegraphics[width=1.0\textwidth]{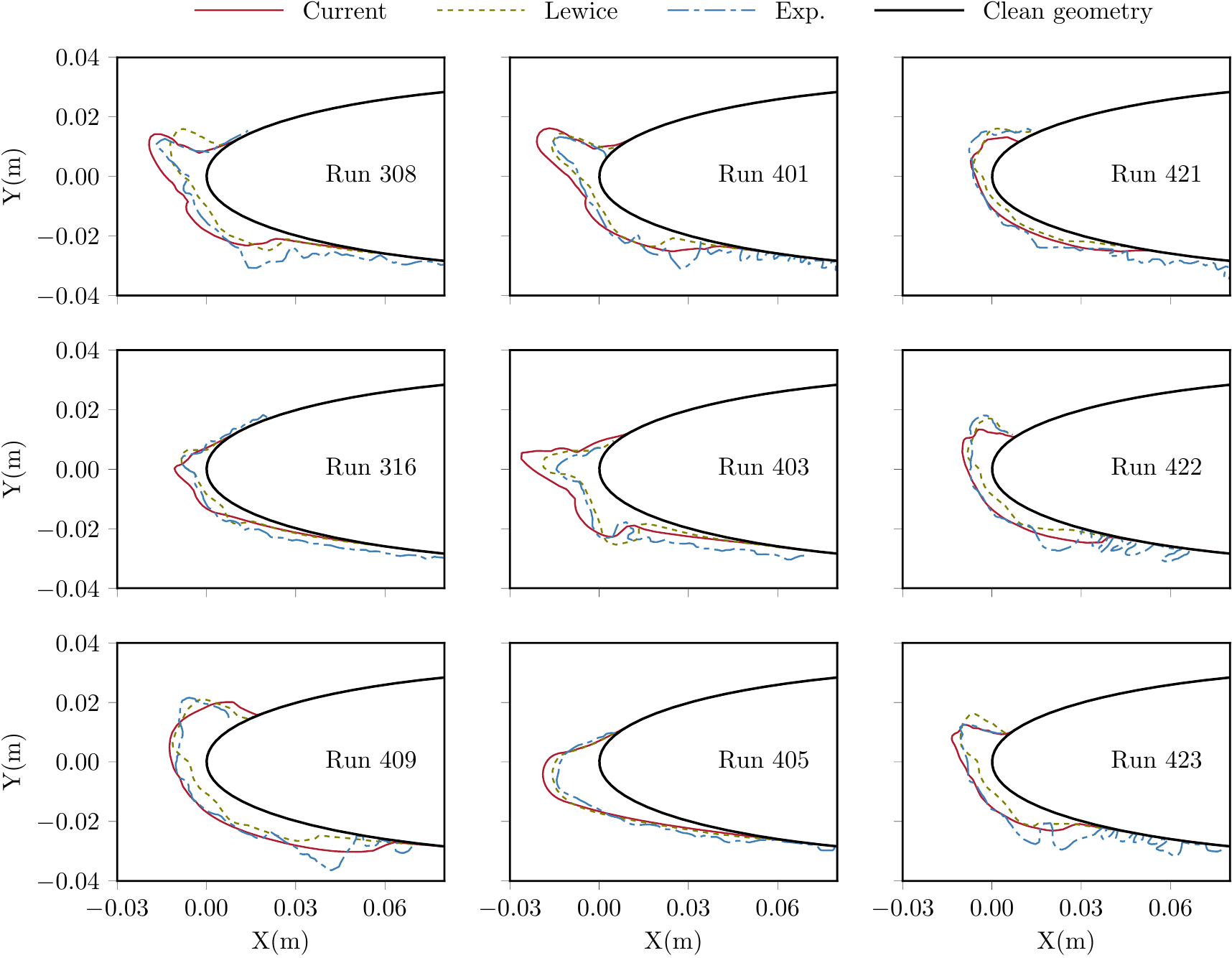}
  \caption{Ice accretion on NACA 0012 airfoil.}
\end{figure*}

Fig.~\ref{fig:naca0012_ice_shapes} illustrates the comparisons between current ice shapes and ice shapes of IRT experiment and LEWICE 2.0.
Note that figures in the same column share the same X axis, while figures in the same row share the same Y axis.
Run 401, run 403 and run 405 (2nd column of Fig.~\ref{fig:naca0012_ice_shapes}) share the same icing conditions except for the temperature, which decreases from 265.37$\rm K$ to 250.37$\rm K$, and the ice shapes present a transition between glaze ice and rime ice.
For run 405, all impinging droplets freeze, thus typical rime ice and smooth streamwise shape form.
For run 403, temperature increases to 262.04$\rm K$. Unfrozen droplets flow downstream and horns form at the upper and lower surfaces near the trailing edge.
With temperature increasing to 265.37$\rm K$, unfrozen droplets were transported downstream further. Therefore the upper horn shifts to downstream, while the lower horn disappear, resulting in a wider range of ice layer.
It is clearly evident that with the increase of the temperature, the ice layer at the stagnation point grows thicker and the position of the ice horns shifts more downstream.
Run 423, run 422 and run 421 (3rd column of Fig.~\ref{fig:naca0012_ice_shapes}) also share the same icing conditions except for the temperature, which increases from 265.07$\rm K$ to 267.37$\rm K$, and all the ice shapes are typical glaze ice.
The impinging limits of these cases are the same, because the trajectories of the droplets are barely affected by the temperature. As mentioned above, with the increase of the temperature, more film flow downstream, and ice tends to spread thinner and farther along the lower surface.

Overall, good agreement among predicted results by current model, experimental results and that of LEWICE is obtained. For few cases, a small discrepancy between predicted and experimental ice shape is observed at the upper horn.
Main reason may lie in the imprecise prediction of air shear and the convective heat transfer coefficient, and further work will be required to solve this problem.

\subsection{Ice accretion on three-dimensional surface}
\label{sec:3D_case}

The experimental results on the GLC-305 swept wing presented by Papadakis\cite{Papadakis_aerodynamic_2003} have been widely used to validate ice accretion models.
The wing section in the streamwise direction is the GLC-305 airfoil and is constant from the root to the tip, and there is a geometry twist of -4$^{\circ}$ from wing root to wing tip.
More details of the swept wing mentioned above refers to \citet{Papadakis_aerodynamic_2003}.
In this paper, two glaze ice cases, CS10 and IS10, are selected to validate current model, and numerical results are compared to experiment and that of LEWICE.
Section comparisons are made at three spanwise locations A, B and C.
The section at station C is located at the wing root and in the streamwise direction.
The sections at stations B and A are taken normal to the wing leading edge at 36.8\% and 73.6\% semispan, respectively.
The corresponding icing conditions are listed in Table~\ref{tb:glc305}.
Similar to the LEWICE3D, the computations are performed using the single step algorithm.

\begin{table}[H]  
  \label{tb:glc305}
  \caption{Icing conditions for GLC-305 swept wing.}
  \begin{ruledtabular}
    \begin{tabular}{lcc}
      Parameters&CS10&IS10\\
      \hline
      MAC, $\rm m$            & 0.4755 & 0.4755  \\
      $U_{\infty}$, $\rm m/s$ & 89.99  & 67.06   \\
      AOA, deg                & 4      & 4       \\
      $T_{\infty}$, $\rm K$   & 261.87 & 269.26  \\
      $P_{\infty}$, $\rm Pa$  & 101300 & 101300  \\
      LWC, $\rm g/m^3$        & 0.68   & 0.65    \\
      MVD, $\rm \mu m$        & 20     & 20      \\
      Time, $\rm min$         & 10     & 10      \\
    \end{tabular}
  \end{ruledtabular}
\end{table}

\begin{figure}[hbt]
    \centering
    \subfigure[Section A.]{
      \label{subfig:glc305_CS10_section_A}
      \includegraphics[width=8.0cm]{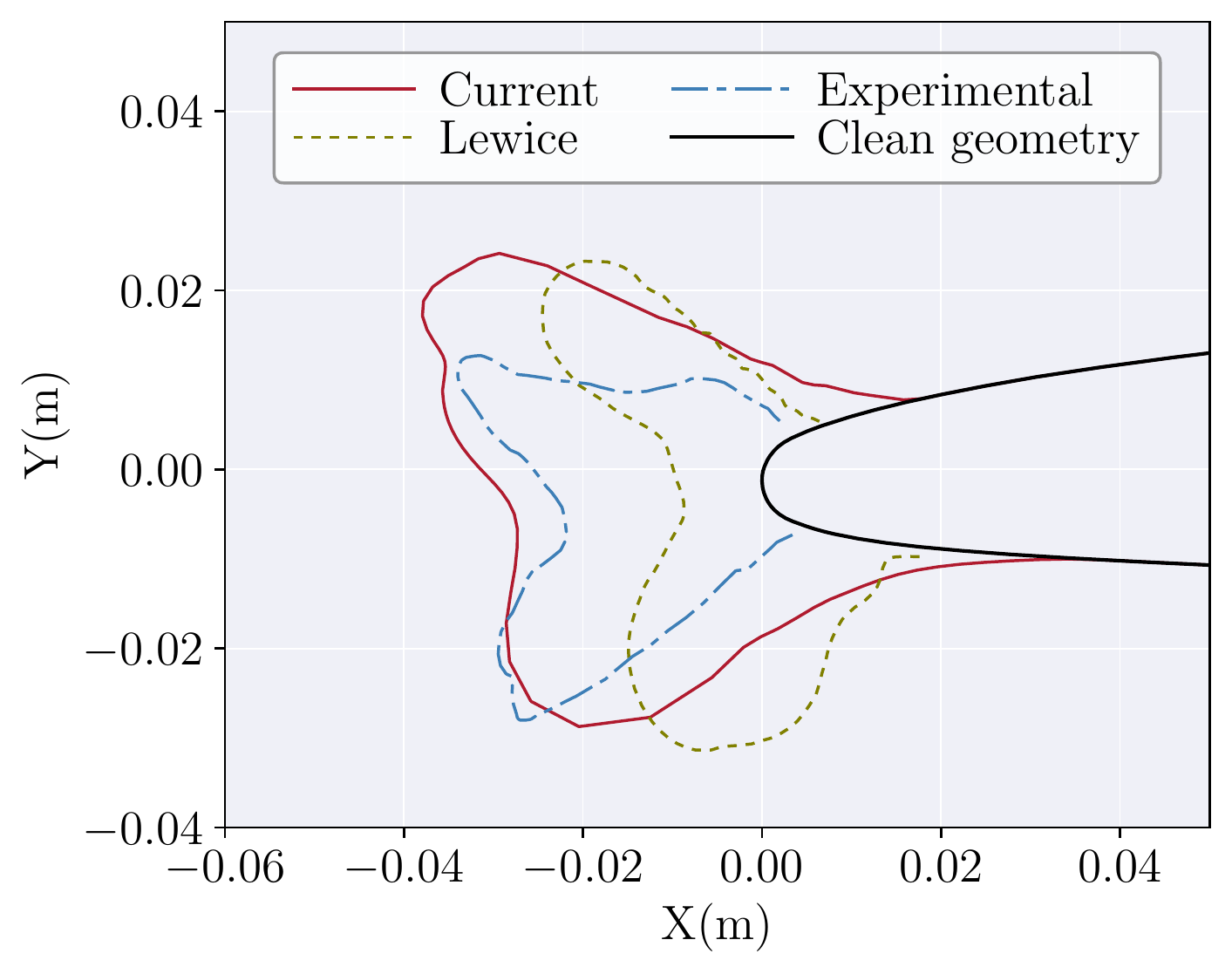}}
    \hspace{0.0cm}
    \subfigure[Section B.]{
      \label{subfig:glc305_CS10_section_B}
      \includegraphics[width=8.0cm]{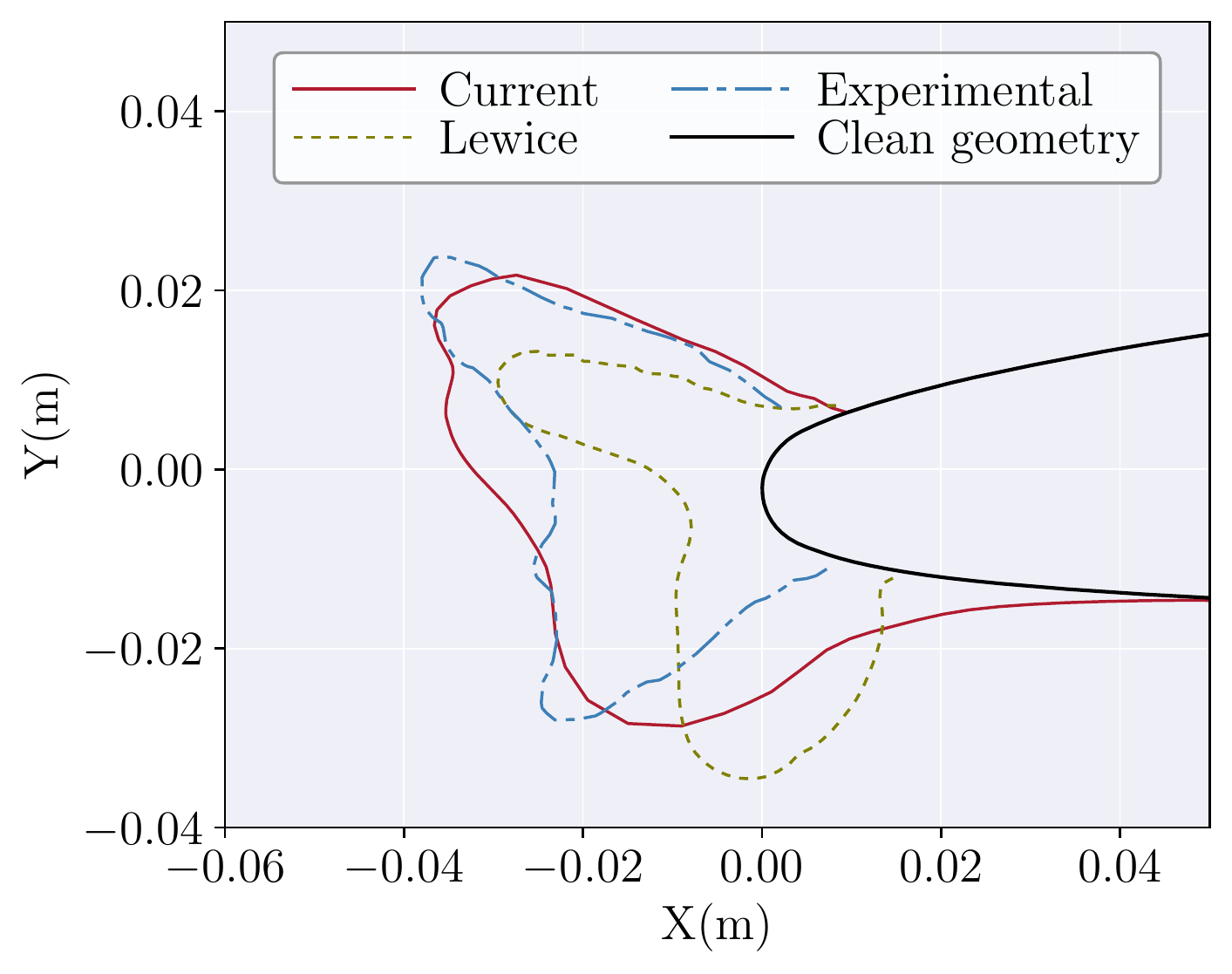}}
    \subfigure[Section C.]{
      \label{subfig:glc305_CS10_section_C}
      \includegraphics[width=8.0cm]{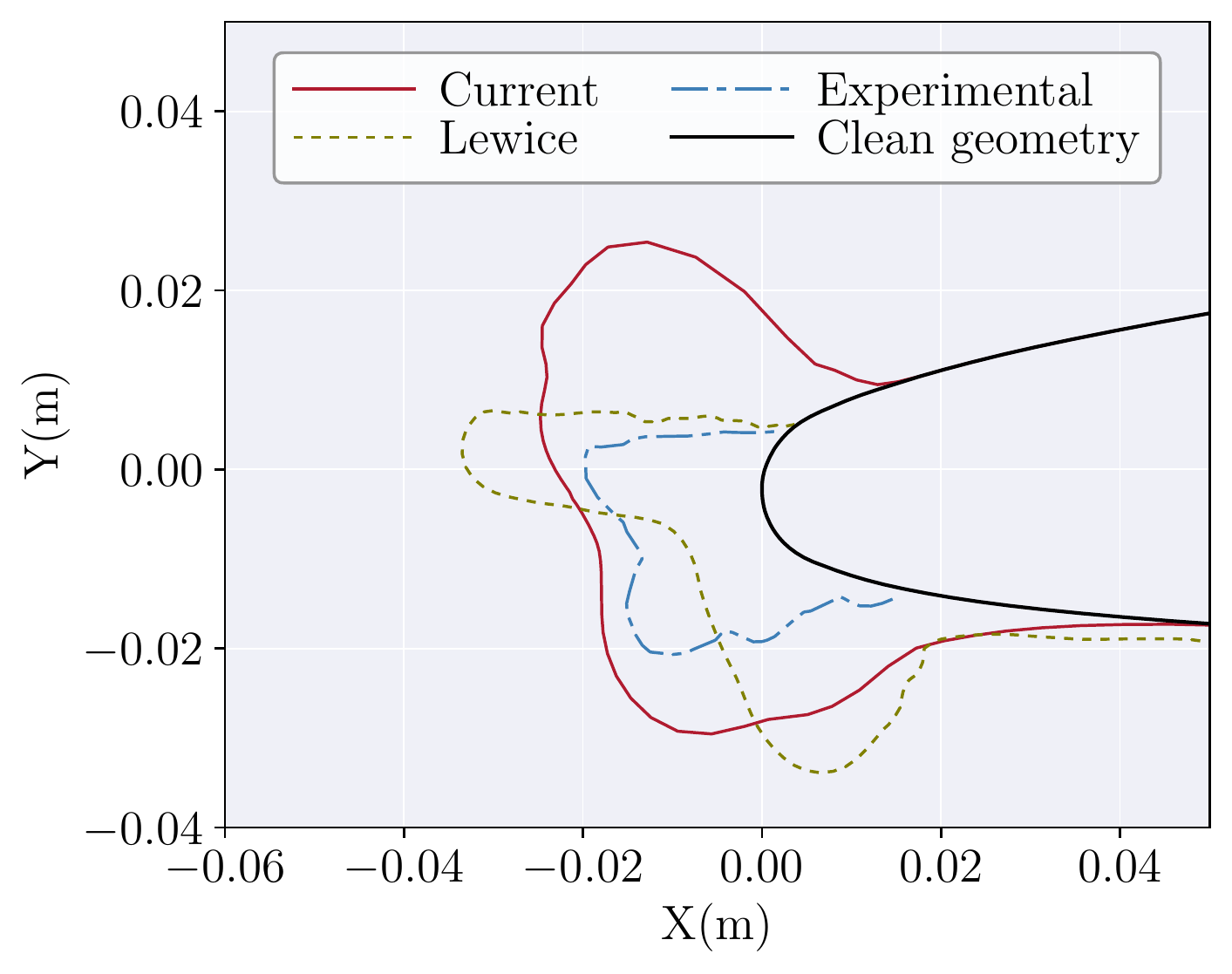}}
    \caption{Ice shapes comparisons of CS10 for GLC-305 swept wing at sections A, B and C.}
    \label{fig:glc305_CS10_section_ABC}
  \end{figure}

\begin{figure}[hbt]
  \centering
  \subfigure[Section A.]{
    \label{subfig:glc305_IS10_section_A}
    \includegraphics[width=8.0cm]{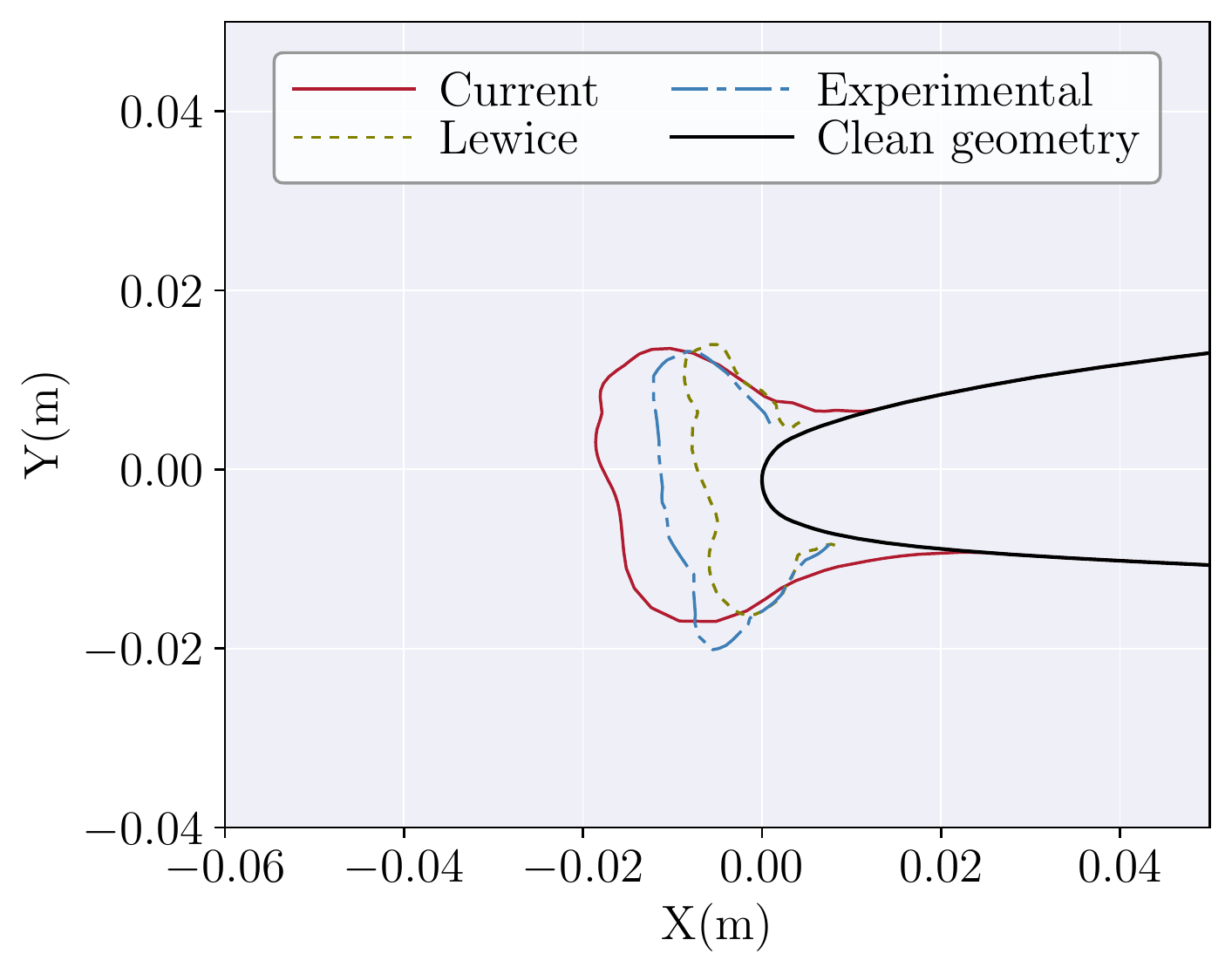}}
  \hspace{0.0cm}
  \subfigure[Section B.]{
    \label{subfig:glc305_IS10_section_B}
    \includegraphics[width=8.0cm]{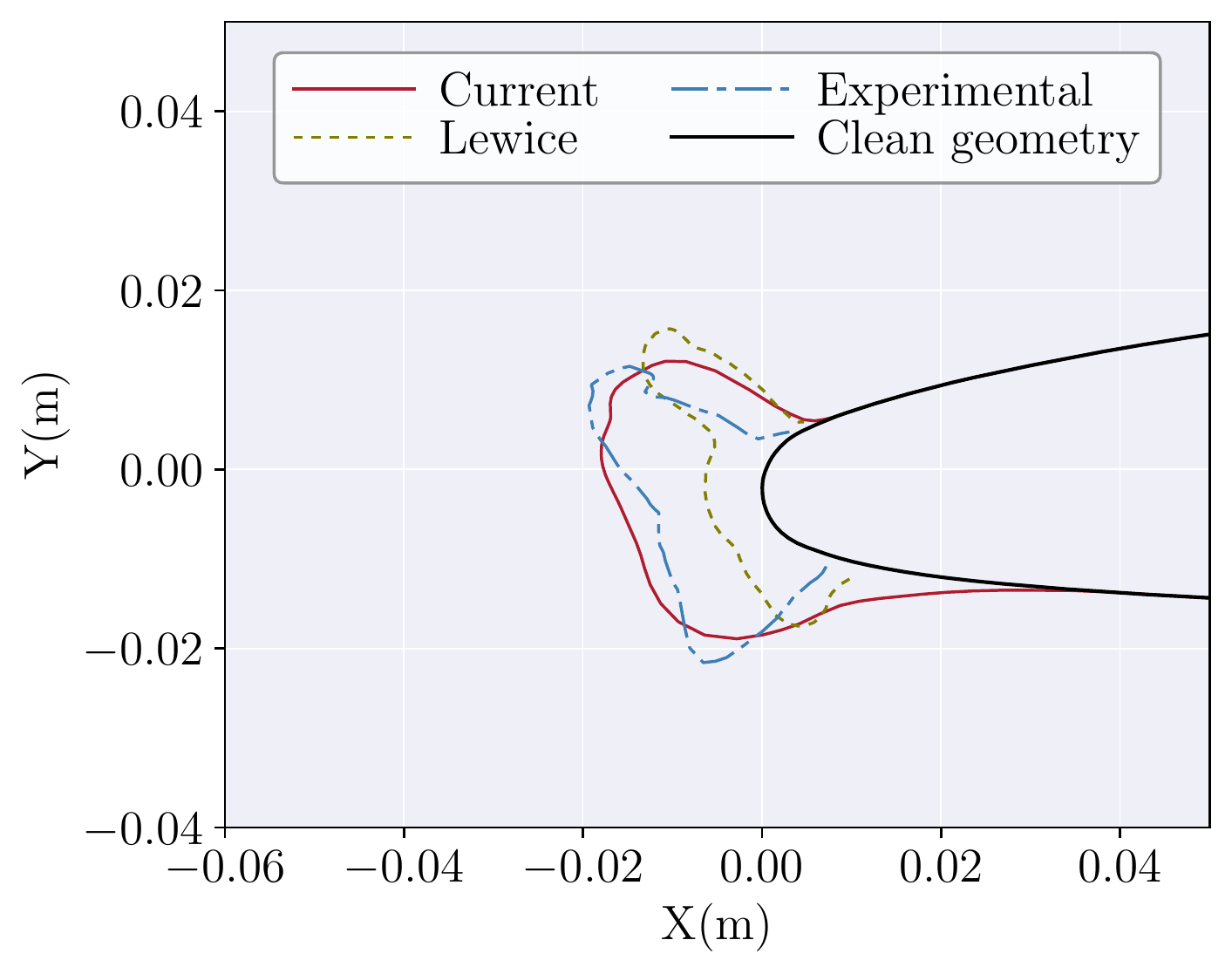}}
  \subfigure[Section C.]{
    \label{subfig:glc305_IS10_section_C}
    \includegraphics[width=8.0cm]{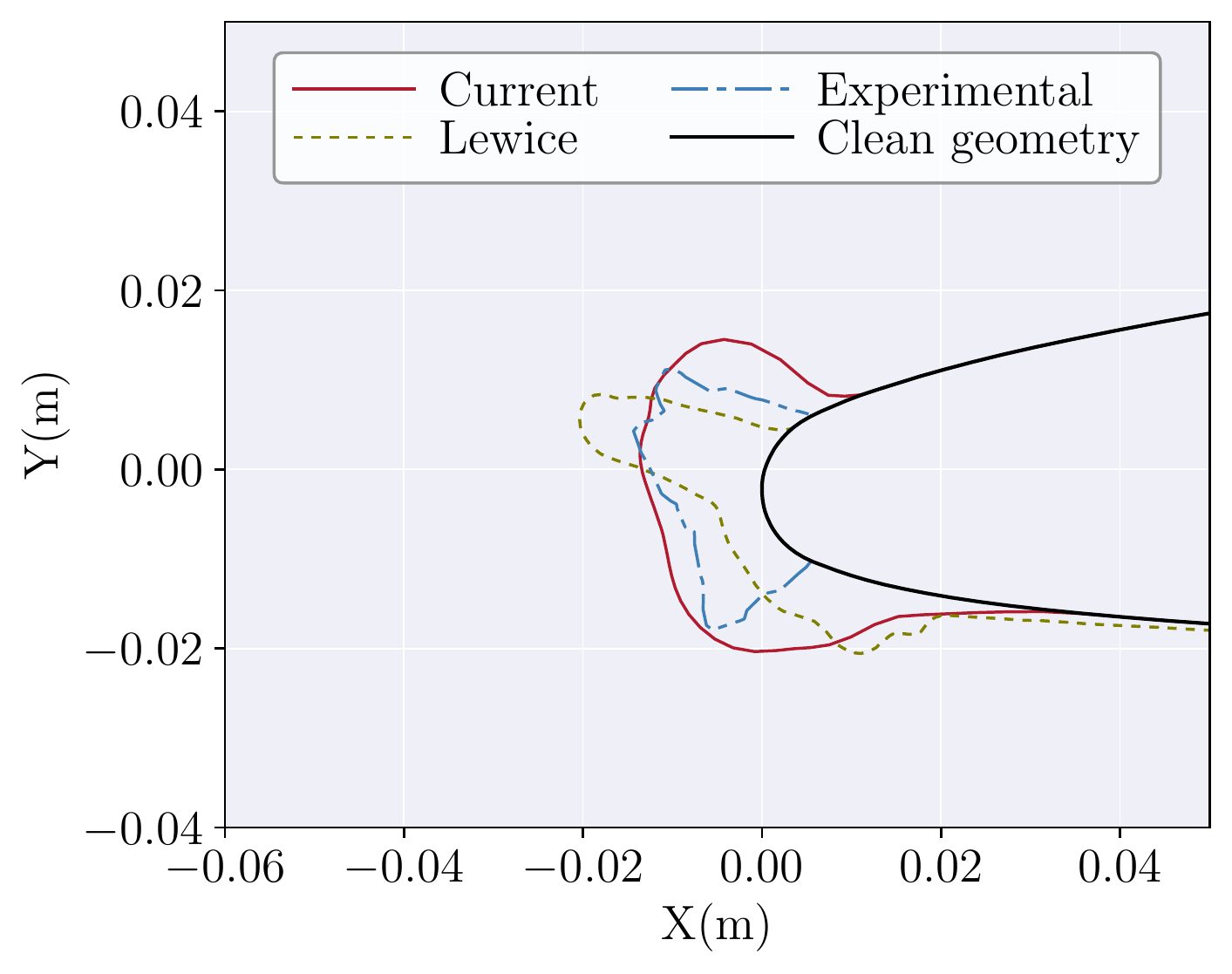}}
  \caption{Ice shapes comparisons of IS10 for GLC-305 swept wing at sections A, B and C.}
  \label{fig:glc305_IS10_section_ABC}
\end{figure}

Fig.~\ref{fig:glc305_CS10_section_ABC} presents the sectional ice accretion shapes of case CS10 at sections A, B and C.
Overall, good agreement in ice accretion predictions is observed for three sections between the numerical results and the experimental results.
The ice thickness around the stagnation point and the ice orientation for all the three sections are in accordance with the experimental data, whereas LEWICE trends to underpredict the ice thickness at the stagnation point for the lack of consideration of the heat conduction through the ice and water layers.
The ice growth direction and the ice thickness at the upper and lower horns are well captured at section A (Fig.~\ref{subfig:glc305_CS10_section_A}) and section B (Fig.~\ref{subfig:glc305_CS10_section_B}).
Among the three sections, section C (Fig.~\ref{subfig:glc305_CS10_section_C}) presents the poorest agreement with the experimental data both for the icing limits and the ice thickness in downstream regions of the stagnation point.
Possible explanations for such differences may be attributed to the fact that in the experiment a fuselage-like body is used to mount the swept wing in the wind tunnel and away from the near-wall flow, while in the numerical simulation a symmetry boundary condition is imposed at the wing root\cite{Papadakis_aerodynamic_2003,cao_extension_2016,liu_three-dimensional_2019}.
Different treatments leads to different flow characteristic near the wing root, which results in the discrepancy in the water collection efficiency and the ice shapes.
Another possible reason may lie in the fact that the ice shape is computed in a single-step method.
Results show that compared to multi-step calculations, single-step calculations significantly overpredicts the ice thickness around the icing limits\cite{verdin_multistep_2009}.
Fig.~\ref{fig:glc305_IS10_section_ABC} presents the sectional ice accretion shapes of case IS10 at sections A, B and C, and the predicted results compared favorably with the experimental results.
Compared to current model, LEWICE still underestimates the ice thickness at the stagnation point.
While one significant difference is observed between predicted results and experiment results.
At section C, the ice thickness around the impinging limits are much thicker than that of experimental results, which is similar with case CS10.

\begin{figure*}[hbt]  
  \centering
  \subfigure[Case CS10.]{
    \label{subfig:beta_htc_span_CS10}
    \includegraphics[width=8.0cm]{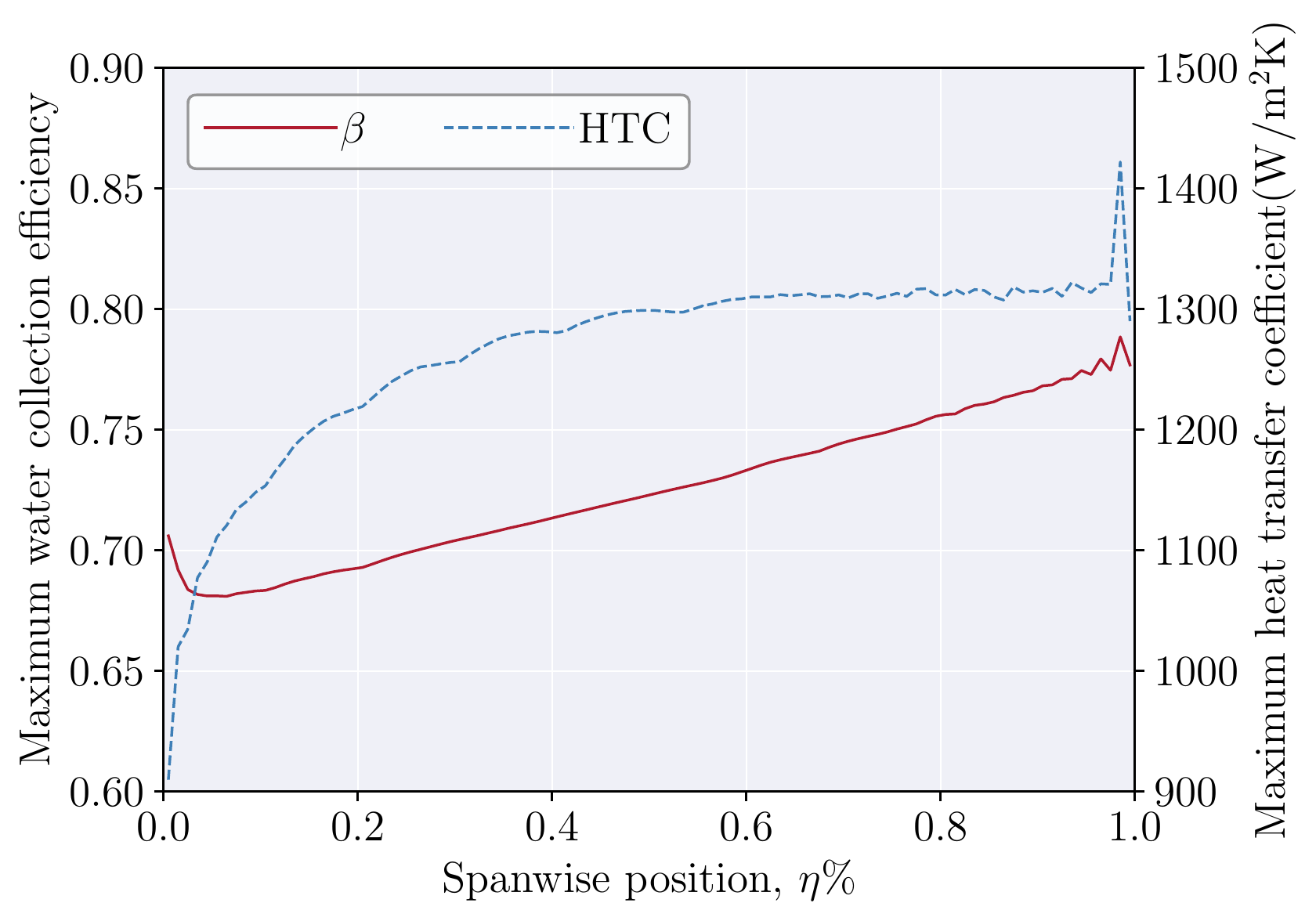}}
  \subfigure[Case IS10.]{
    \label{subfig:betta_htc_span_IS10}
    \includegraphics[width=8.0cm]{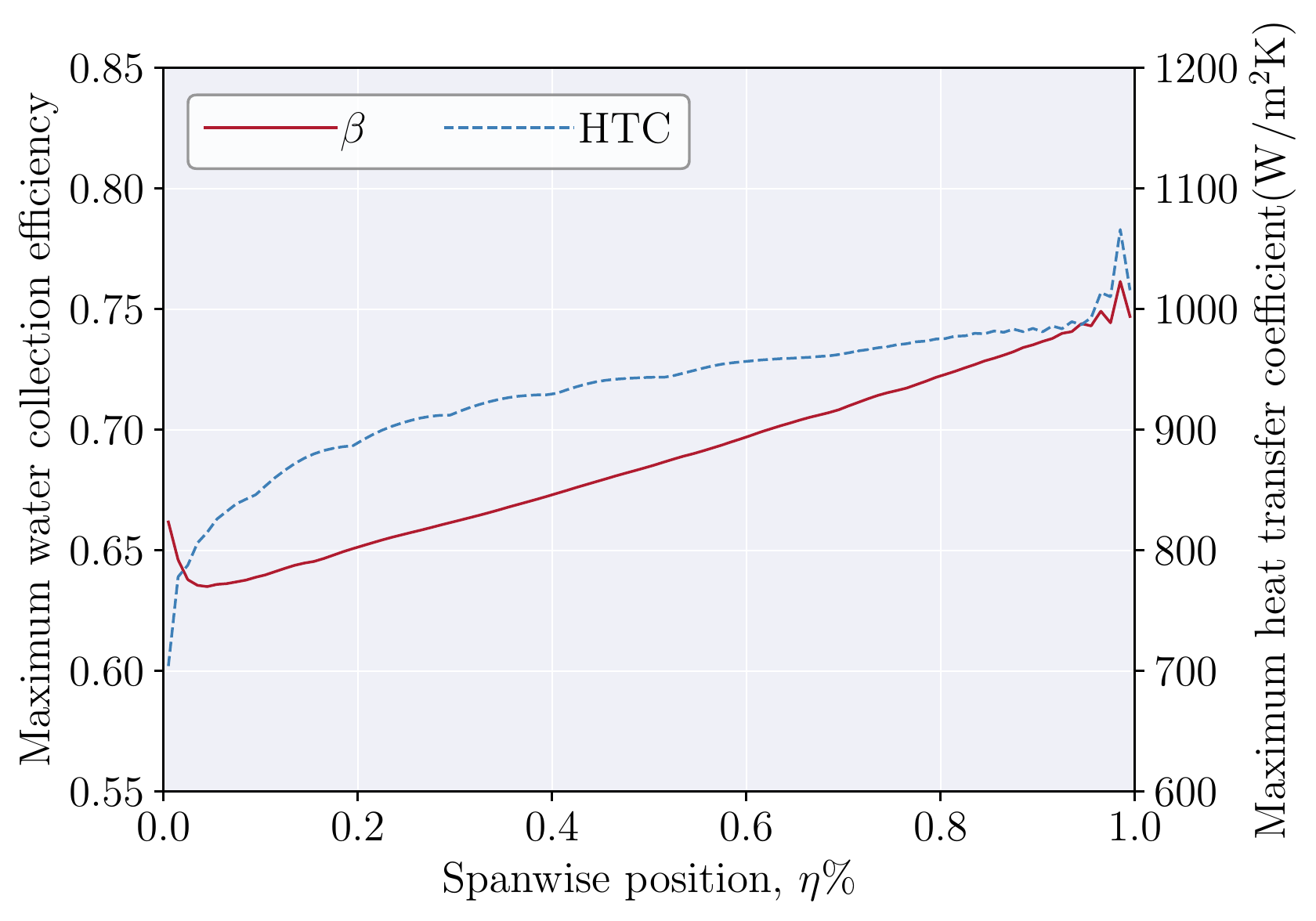}}
  \caption{Water collection efficiency and convective heat transfer coefficient comparisons across the span for case CS10 and case IS10.}
  \label{fig:betta_htc_span}
\end{figure*}

\begin{figure*}[hbt]  
  \centering
  \subfigure[Case CS10.]{
    \label{subfig:film_thickness_span_CS10}
    \includegraphics[width=8.0cm]{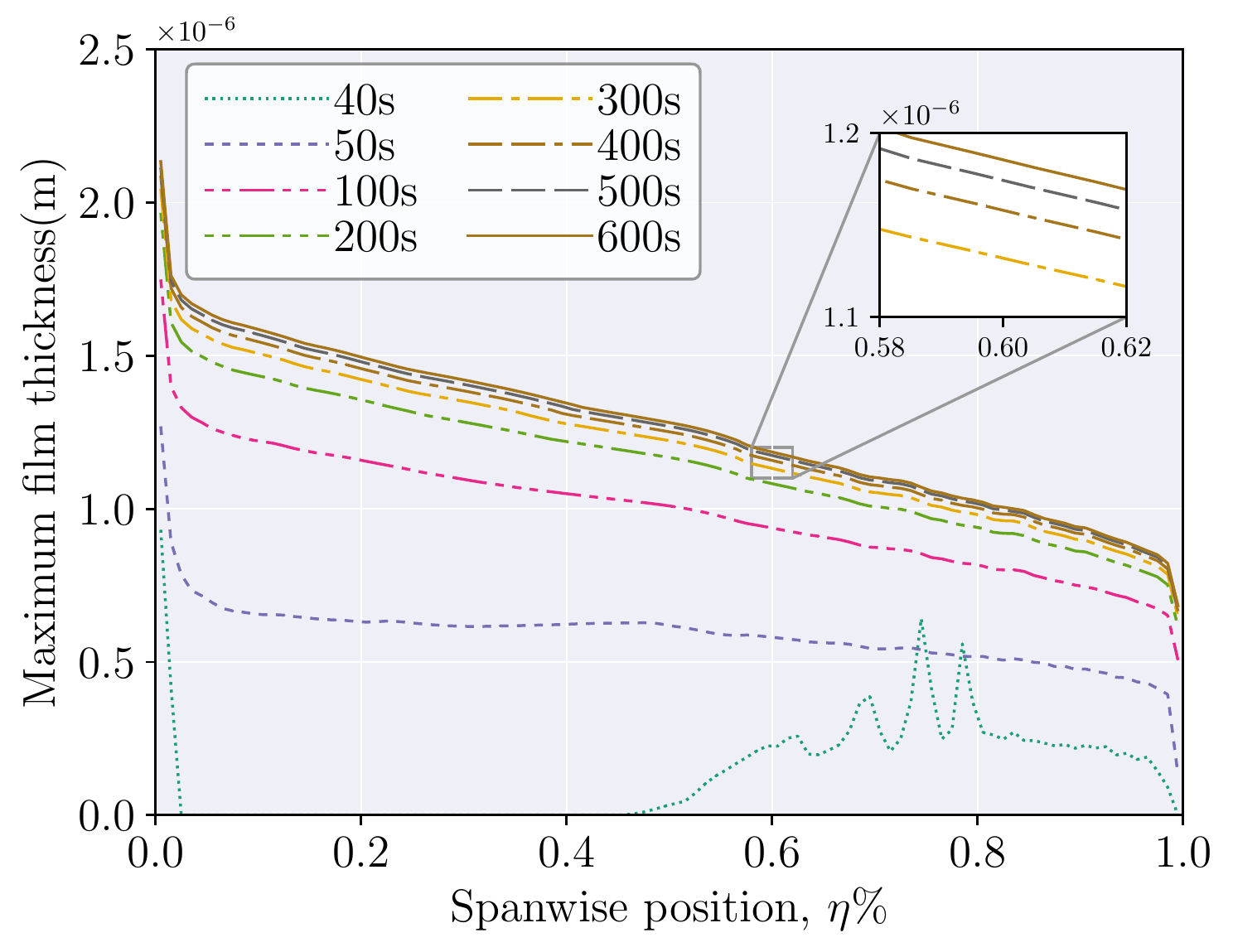}}
  \subfigure[Case IS10.]{
    \label{subfig:film_thickness_span_IS10}
    \includegraphics[width=8.0cm]{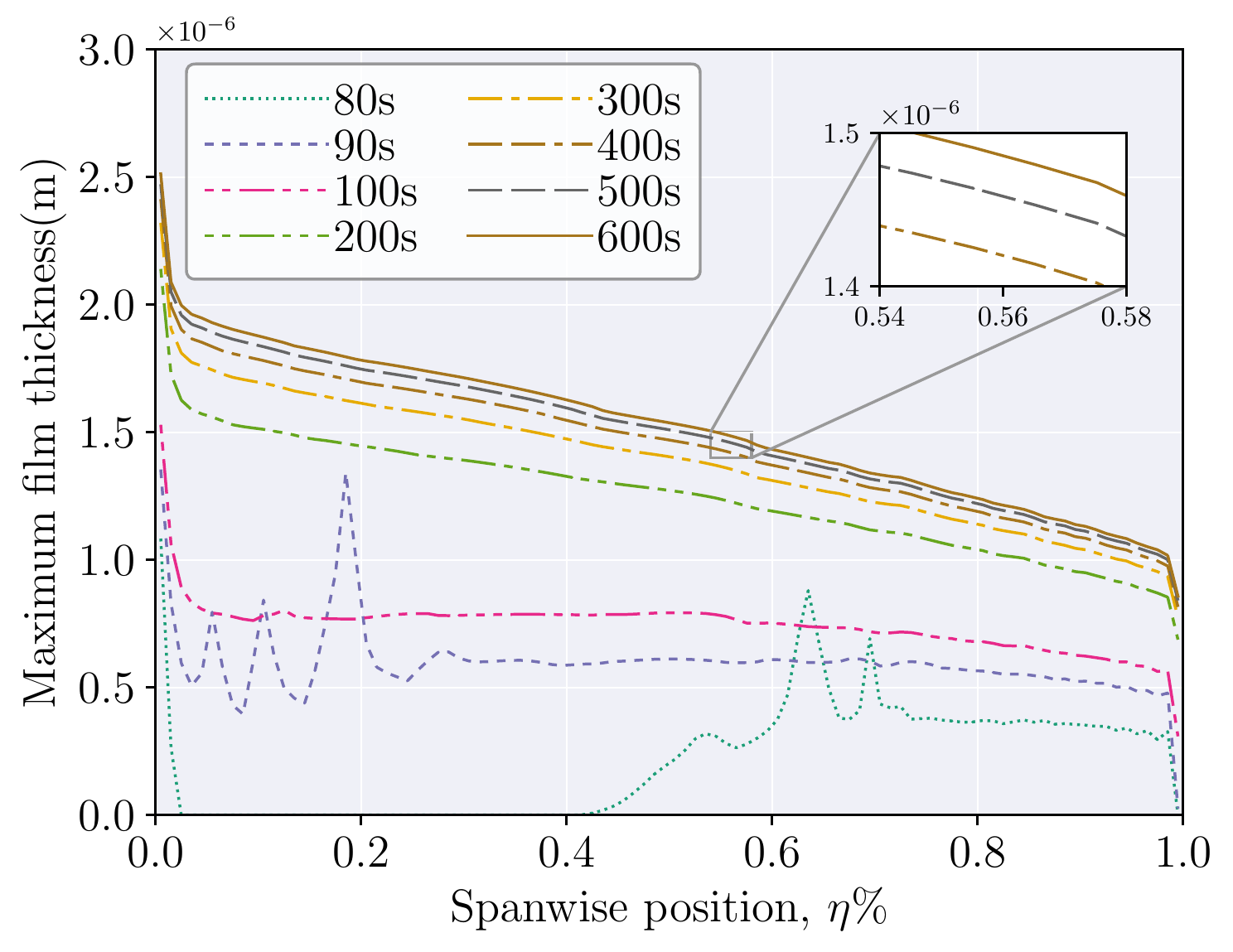}}
  \caption{Film thickness comparison across the span for case CS10 and case IS10.}
  \label{fig:film_thickness_span}
\end{figure*}

\begin{figure*}[hbt]  
  \centering
  \subfigure[Case CS10.]{
    \label{subfig:ice_thickness_span_CS10}
    \includegraphics[width=8.0cm]{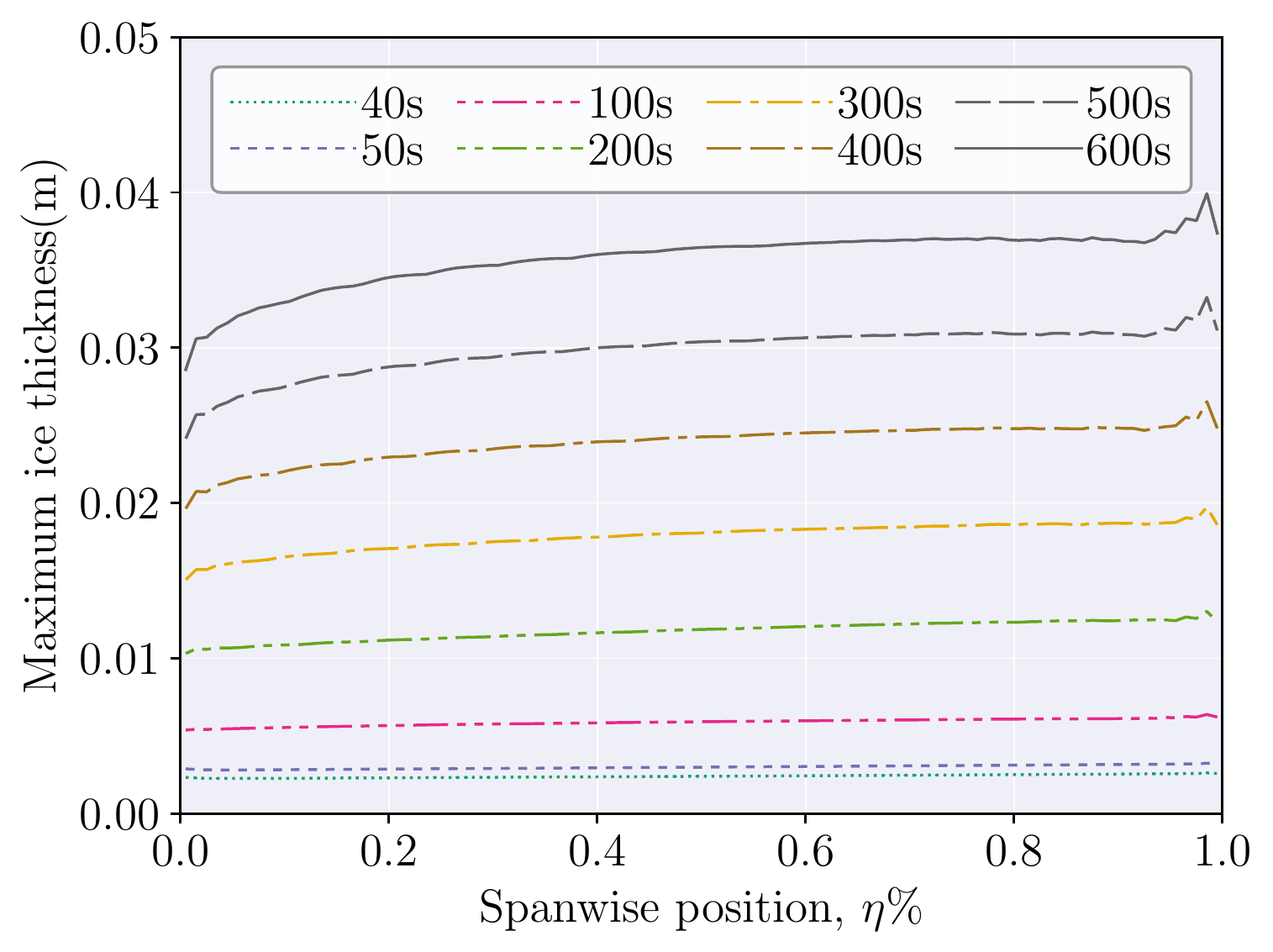}}
  \subfigure[Case IS10.]{
    \label{subfig:ice_thickness_span_IS10}
    \includegraphics[width=8.0cm]{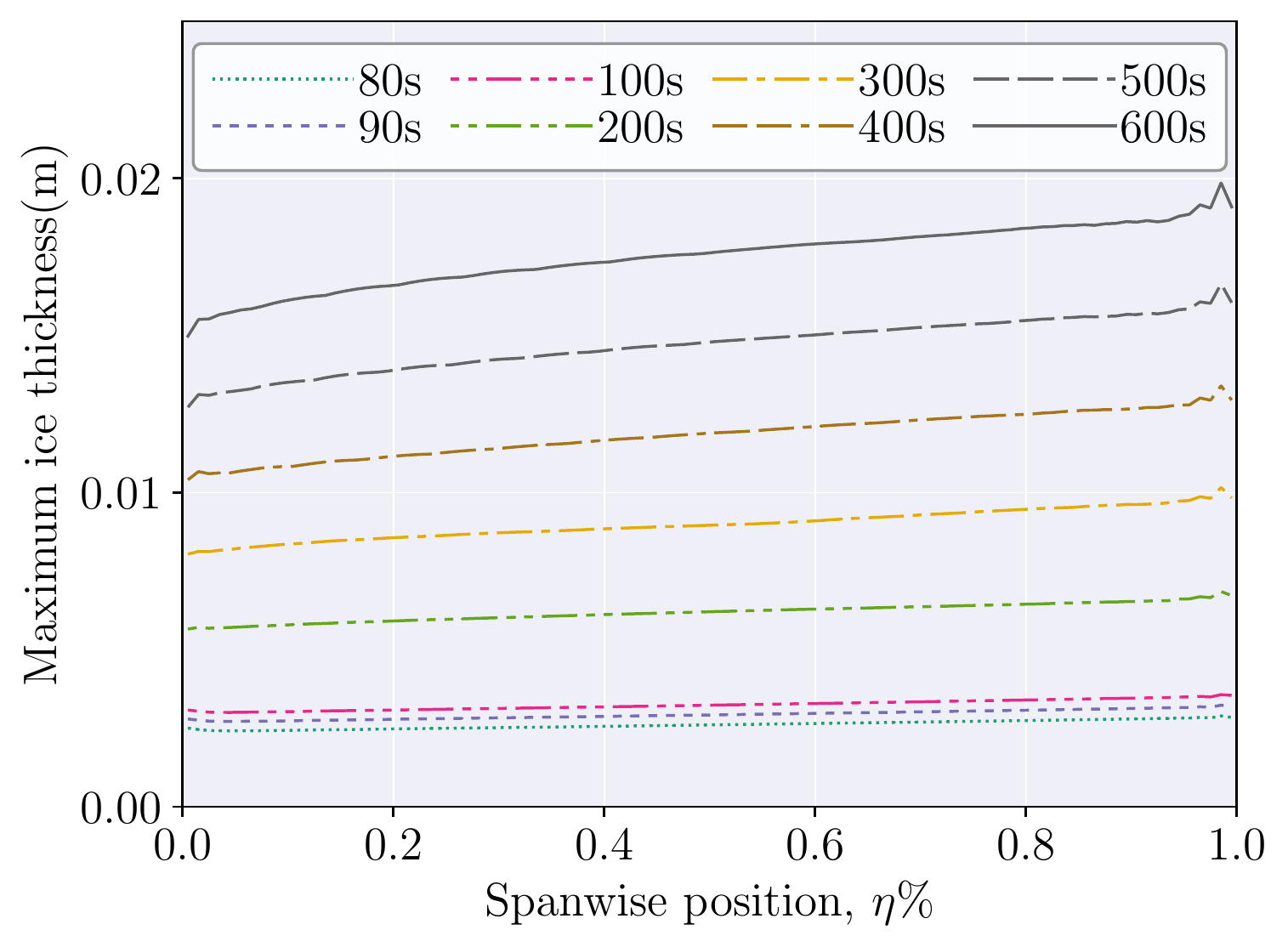}}
  \caption{Ice thickness comparison across the span for case CS10 and case IS10.}
  \label{fig:ice_thickness_span}
\end{figure*}

Fig.~\ref{fig:betta_htc_span} shows the maximum water collection efficiency $\beta$ and convective heat transfer coefficient HTC in the streamwise direction across all the semispan positions, while Fig.~\ref{fig:film_thickness_span} and Fig.~\ref{fig:ice_thickness_span} illustrate the evolution of the maximum film and ice thickness in the streamwise direction across all the semispan positions.
Note that the spanwise position is normalized by the semispan length.
CS10 and IS10 are typical glaze ice case, therefore similar trends of $\beta$, HTC, film thickness and ice thickness are observed for these two cases.
Compared to CS10, less flow velocity of IS10 leads to, on the whole, less water collection efficiency, and finally less ice accretion volume.

\begin{figure*}[hbt]
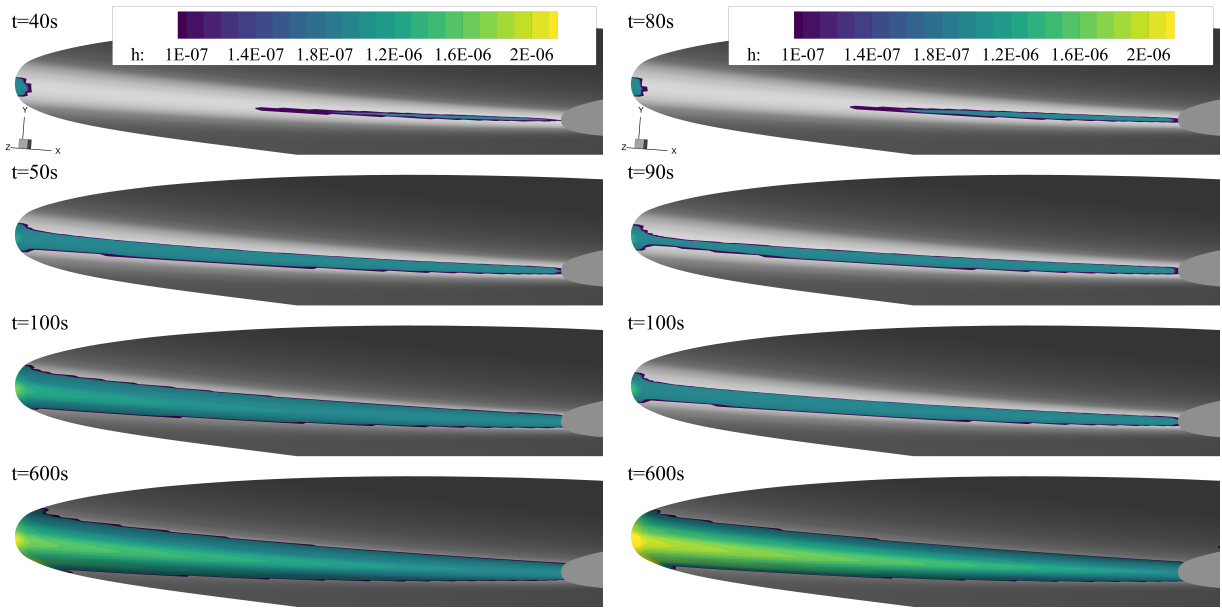
  
  \centering
  \subfigure[Case CS10.]{
    \label{subfig:film_thickness_evolution_CS10}
    \includegraphics[width=8.0cm]{CS10_film_evolution}}
  \subfigure[Case IS10.]{
    \label{subfig:film_thickness_evolution_IS10}
    \includegraphics[width=8.0cm]{IS10_film_evolution}}
  \caption{Film thickness evolution for case CS10 and case IS10.}
  \label{fig:film_thickness_evolution}
\end{figure*}

As shown in Fig.~\ref{fig:betta_htc_span}, $\beta$ and HTC increase from the wing root to the wing tip, which indicates that the wing tip tends to collect more supercooled water and the supercooled water are more likely to freeze near the wing tip.
This leads to the decrease of the film thickness in the outward spanwise direction, as shown in Fig.~\ref{fig:film_thickness_span} and Fig.~\ref{fig:film_thickness_evolution}.
As we can see in these two figures, film first appears at the wing root and near the wing tip after about 40$\mathrm{s}$.
Then the film spreads downstream, and the overall film thickness increases gradually over time, which means that more unfrozen water run back over the wing surface and freeze in downstream regions.
After about 300$\mathrm{s}$, the overall film thickness trends to reach some constants, and the film flow becomes stable, which indicates that the mass of water entering the control cell is nearly equal to that leaving the control cell.
Furthermore, film thickness at the wing root is significantly greater than that at any other regions because of the smaller HTC at the wing root.
As a result, the predicted ice horn at section C is more distinct than that at section B and section A.

\section{Conclusion}
\label{sec:conclusion}

Based on the finite area method, we derive the governing equations for film flow and ice accretion process on arbitrary curved surfaces.
The governing equations are expressed in Cartesian coordinate system therefore it is easy to be coupled to existing FVM flow solver Exstream.
The film flow flow is dominated by shear stress, pressure gradient and gravity.
The velocity field is expressed in Cartesian coordinate system. 
Besides, the velocity profile in the film is approximated as a polynomial function with respect to the film thickness and finally the depth-averaged velocity of the film can be expressed as a function of the film thickness and the shear stress acting on the film.
Numerical simulations are conducted on airfoils and the wing, and current film and ice accretion model are verified to be reasonable by calculating the flow behavior of the runback water and the shape feature of the ice shapes.
Numerical results show that, for some glaze ice conditions, as temperature increases, ice tends to extend higher on the upper surface and spread thinner and farther along the lower surface.

\begin{acknowledgments}
This work was supported by the National Natural Science Foundation of China (No. 11902271 and No. 91952203), the Fundamental Research Funds for the Central Universities of China (No. G2020KY05101),  the Foundation of National Key Laboratory (No. 6142201190303), and the 111 Project of China (B17037). The authors acknowledge the computational resources provided by the Tianhe-2 supercomputer of the National Supercomputer Center in Guangzhou (NSCC-GZ).
\end{acknowledgments}

\appendix

\section{Moment conservation}
\label{app:moment_conservation}

The first term of the left-hand-side of  Eq.~\eqref{eq:moment_conservation_FVM} can be decomposed as
\begin{equation}
  \label{eq:pressure_eq_sum_of_three_terms}
  \int_S p \bm{n} {\rm d}S
  = \int_{S_{io}} p \bm{n}_{io} {\rm d}S 
  + \int_{S_b}    p_b \bm{n}_{b}  {\rm d}S 
  + \int_{S_{fs}} p_{fs} \bm{n}_{fs} {\rm d}S 
\end{equation}
where $p_b$ and $p_{fs}$ are the pressure at the bottom and the surface of the film respectively.
Here a simple linear pressure profile in the film layer is assumed, then the pressure along the film thickness is $p(\zeta) = p_b + \dfrac{\zeta}{h}(p_{fs} - p_b)$.
Applying the depth-integration to the first term of the right-hand-side of Eq.~\eqref{eq:pressure_eq_sum_of_three_terms} we get
\begin{equation}
  \begin{split}
    \int_{S_{io}} p \bm{n}_{io} {\rm d}S
    &= \oint_{L_{io}} \bm{n}_{io} \int_0^h p {\rm d}\zeta {\rm d}L \\
    &= \oint_{L_{io}} \bm{n}_{io} h \bar{p} {\rm d}L
  \end{split}
\end{equation}
where the depth-averaged pressure $\bar{p} = \dfrac{1}{2}\left( p_{fs} + p_b \right)$.
At last Eq.~\eqref{eq:pressure_eq_sum_of_three_terms} gives
\begin{equation}
  \label{eq:pressure_term_simplified}
  \int_S p \bm{n} {\rm d}S
  = \oint_{L_{io}} \bm{n}_{io} h \bar{p} {\rm d}L 
  + \int_{S_b} (p_b-p_a) \bm{n}_{b}  {\rm d}S 
\end{equation}

For the second term of the left-hand-side of Eq.~\eqref{eq:moment_conservation_FVM}, by performing the depth-integration the body force induced by the gravity acceleration force can be easily obtained as
\begin{equation}
  \label{eq:gravity_term_simplified}
  \int_V \rho_w \bm{g} {\rm d}V
  = \int_{S_b} h \rho_w \bm{g} {\rm d}S
\end{equation}

The third term of the left-hand-side of Eq.~\eqref{eq:moment_conservation_FVM} can be split into
\begin{equation}
  \label{eq:div_eq_three_terms}
  \begin{split}
    \int_V \mu_w \nabla^2 \bm{u} {\rm d}V
    &= \oint_S \mu_w \nabla \bm{u} \cdot \bm{n} {\rm d}S \\
    &= \int_{S_{io}} \mu_w \nabla \bm{u} \cdot \bm{n}_{io} {\rm d}S
     + \int_{S_{b}}  \mu_w \nabla \bm{u}_b \cdot \bm{n}_{b} {\rm d}S \\
     &+ \int_{S_{fs}} \mu_w \nabla \bm{u}_{fs} \cdot \bm{n}_{fs} {\rm d}S
  \end{split}
\end{equation}
The first term of the right-hand-side of Eq.~\eqref{eq:div_eq_three_terms} can be written as
\begin{equation}
  \int_{S_{io}} \mu_w \nabla \bm{u} \cdot \bm{n}_{io} {\rm d}S
  = \oint_{L_{io}} \mu_w \int_0^h \nabla \bm{u} \cdot \bm{n}_{io} {\rm d}\zeta {\rm d}L
\end{equation}
According to the shear continuity at the water-air interface, $\mu_w \nabla \bm{u}_{fs} \cdot \bm{n}_{b} = \bm{\tau}_a$, the third term can be written as
\begin{equation}
  \begin{split}
    \int_{S_{fs}} \mu_w \nabla \bm{u}_{fs} \cdot \bm{n}_{fs} {\rm d}S
    &\approx - \int_{S_{b}} \mu_w \nabla \bm{u}_{fs} \cdot \bm{n}_{b} {\rm d}S \\
    &= - \int_{S_{b}} \bm{\tau}_a {\rm d}S
  \end{split}
\end{equation}

At the water-substrate/ice interface, we let $\mu_w \nabla \bm{u}_b \cdot \bm{n}_b = \bm{\tau}_b$. Finally Eq.~\eqref{eq:div_eq_three_terms} is
\begin{equation}
  \label{eq:div_eq_three_terms_simplified}
  \begin{split}
    \int_V \mu_w \nabla^2 \bm{u} {\rm d}V
    &= \oint_S \mu_w \nabla \bm{u} \cdot \bm{n} {\rm d}S \\
    &= \oint_{L_{io}} \mu_w \int_0^h \nabla \bm{u} \cdot \bm{n}_{io} {\rm d}\zeta {\rm d}L \\
     &+ \int_{S_{b}}  \bm{\tau}_b {\rm d}S - \int_{S_{b}} \bm{\tau}_a {\rm d}S
  \end{split}
\end{equation}

Finally, substituting Eq.~\eqref{eq:pressure_term_simplified}, Eq.~\eqref{eq:gravity_term_simplified} and Eq.~\eqref{eq:div_eq_three_terms_simplified} into Eq.~\eqref{eq:moment_conservation_FVM} we obtain the moment conservation equation Eq.~\eqref{eq:moment_conservation_FAM}.

To solve the pressure at the bottom surface, we multiply Eq.~\eqref{eq:differential_moment_equation} with the normal vector $\bm{n}_b$:
\begin{equation}
  \label{eq:normal_differential_moment_equation}
  \begin{split}
    &- \bm{n}_b \cdot \hat{\nabla} \left( h\bar{p} \right) - (p_b-p_a) \bm{n}_b \cdot \bm{n}_b \\
    &+ h \rho_w \bm{n}_b \cdot \bm{g}
    + \bm{n}_b \cdot (\bm{\tau}_b - \bm{\tau}_a)
    = 0
  \end{split}
\end{equation}
The first and the last terms at the left-hand-side of Eq.~\eqref{eq:normal_differential_moment_equation} are approximately 0, and then the equation along the normal direction can be deduced to
\begin{equation}
  \label{eq:app_normal_moment_equation}
  -(p_b - p_a) + h \rho_w g_n = 0
\end{equation}
and thus we obtain
\begin{equation}
  p_b = p_a + h \rho_w g_n
\end{equation}
and further
\begin{equation}
  \bar{p} = p_a + \frac{1}{2} h \rho_w g_n
\end{equation}

To evaluate the surface tangential momentum equation, we multiply Eq.~\eqref{eq:app_normal_moment_equation} with $\bm{n}_b$ again and it yields
\begin{equation}
  -(p_b - p_a) \bm{n}_b + h \rho_w \bm{g}_n = 0
\end{equation}
where $\bm{g}_n = \bm{n}_b \cdot \left( \bm{n}_b \cdot \bm{g} \right)$. Subtracting the equation above with Eq.~\eqref{eq:differential_moment_equation} yields
\begin{equation}
  \label{eq:normal_normal_moment_equation}
  - \hat{\nabla}_s \left( h \bar{p} \right)
  + h \rho_w \bm{g}_s
  + (\bm{\tau}_b - \bm{\tau}_a)
  = 0
\end{equation}

The depth-averaged velocity of the film is
\begin{equation}
  \label{eq:app_vector_depth_averaged_velocity_coeff}
  \bar{\bm{u}} = \frac{\bm{a}}{3} h^2 + \frac{\bm{b}}{2} h + \bm{c}
\end{equation}
While the gradient of velocity is
\begin{equation}
  \frac{\partial \bm{u}}{\partial \zeta}
   = 2 \bm{a} \zeta + \bm{b}.
\end{equation}
The no-slip boundary condition at the water-substrate interface gives $\bm{c} = 0$ directly.
Applying the shear stress boundary condition~\eqref{eq:shear_stress_continuty} yields
\begin{equation}
  \bm{a} = \frac{\bm{\tau}_a - \bm{\tau}_b}{2 h \mu_w}
\end{equation}
While according to the defination of $\bm{\tau}_b$ we can easily obtain
\begin{equation}
  \bm{b} = \frac{\bm{\tau}_b}{\mu_w}
\end{equation}
At last, substituting the coefficients above into Eq.~\eqref{eq:app_vector_depth_averaged_velocity_coeff} yields
\begin{equation}
  \bar{\bm{u}}
   = \frac{\left( \bm{\tau}_a + 2 \bm{\tau}_b \right)h}{6 \mu_w}
\end{equation}

\nocite{*}
\bibliography{ref}

\end{document}